\documentclass[reprint,superscriptaddress,amsmath,amssymb,aps]{revtex4-2}
\usepackage[colorlinks=true,linkcolor=red,citecolor=blue,urlcolor=blue]{hyperref}
\usepackage{graphicx}
\usepackage{orcidlink}
\usepackage{xcolor}
\usepackage{float}
\begin{document}

\title{Constraining the High-Density Equation of State with Present and Future NICER Observations Using Physics-Informed Regularized Machine Learning}

\author{Utkarsh Atul Deshmukh \orcidlink{0009-0009-7002-1054}}
\email{utkarshd21@iiserb.ac.in}
\affiliation{Indian Institute of Science Education and Research Bhopal, Bhopal 462066, India}

\author{Asim Kumar Saha \orcidlink{0009-0000-8375-4833}}
\email{asim21@iiserb.ac.in}
\affiliation{Indian Institute of Science Education and Research Bhopal, Bhopal 462066, India}

\author{Ritam Mallick \orcidlink{0000-0003-2943-6388}}
\email{mallick@iiserb.ac.in}
\affiliation{Indian Institute of Science Education and Research Bhopal, Bhopal 462066, India}

\date{March 2026}

\begin{abstract}
The precise mass and radius measurements of neutron stars by NICER have significantly advanced our ability to constrain the properties of matter at supranuclear densities. In this work, we develop a physics-informed regularized conditional Invertible Neural Network (cINN) that bijectively maps mass--radius posterior distributions directly onto the corresponding central energy density and pressure, eliminating the need for explicit high-dimensional parameter sampling. The physics-informed regularisation guarantees that all inferred solutions satisfy causality and thermodynamic stability, ensuring physically consistent predictions without explicit forward modelling. We demonstrate that the framework accurately reconstructs central EoS posteriors for NICER-like observations while preserving the mapping between macroscopic stellar observables and the microscopic properties of dense matter. Exploiting the computational efficiency of the cINN, we perform a systematic optimisation study of 62,400 simulated mass--radius observations to identify the most informative targets for constraining the high-density EoS. We find that the constraining power depends strongly on the location of the observation in the mass--radius plane, with an optimal strategy that alternates between compact high-mass stars and extended intermediate-mass stars, reducing the uncertainty in the inferred EoS by up to $\sim 9\%-10\%$ relative to the current NICER baseline. These results establish physics-informed invertible neural networks as a powerful framework for rapid, physically consistent inference of dense-matter properties from present and future multi-messenger observations.

\end{abstract}

\maketitle

\section{Introduction}

Understanding the Equation of State (EoS) of dense nuclear matter remains one of the central challenges in modern astrophysics \cite{Baym_2018, Lattimer_2012}. Simulating such extreme conditions, specifically cold, highly neutron-rich matter at several times the nuclear saturation density, is still not possible in terrestrial heavy-ion colliders. Therefore, Neutron Stars (NS) serve as our only natural laboratories for probing this regime \citep{ Shuryak:1980tp,ShapiroTeukolsky1983, Glendenning2000, LattimerPrakash2007, OzelFreire2016, Oertel2017}. Supplying a theoretical EoS into the Tolman-Oppenheimer-Volkoff (TOV) equations, one obtains the macroscopic observables of the star, namely its mass and radius\citep{Oppenheimer1939,Tolman1939}.

However, due to current technological and observational constraints, the measurement error in the radius of a neutron star is considerable. Instead, current multi-messenger detections yield broad Mass-Radius ($M$-$R$) posterior regions, characterised by varying levels of statistical confidence. Despite the imprecise observations, these posteriors are incredibly valuable, allowing us to test the fundamental nature of the strong interaction under conditions entirely inaccessible to lattice Quantum Chromodynamics (QCD) and Earth-based experiments \cite{Fukushima_2010, Watts_2016, Capano_2020, Dietrich_2020}. Employing an ensemble of parameterised EoS (as discussed in Section \ref{sec:data}) into the TOV equations, one generates a family of $M$-$R$ curves. A subset of these curves that simultaneously satisfy current NS observational posteriors allows us to isolate their underlying physics, thereby defining a constrained band of physically viable models \cite{Cromartie_2020, Fonseca_2021,Imam:2024gfh, ecker_2023, Greif:2018njt, Rezzolla_2018, Saha:2024swd}. This band represents the allowed ranges of pressure ($p$) and energy density ($\epsilon$) within a calculated uncertainty. Ultimately, reducing uncertainty in the EoS directly affects the understanding of dense matter\cite{Imam:2021dbe, Biswas_2022, Roy:2024sjx, Annala:2021gom, Tewari:2024qit, Saha:2024swd, Imam:2025lut, Verma:2025dez, Albino:2025puc, Cartaxo:2025jpi, Saha:2025don, Annala:2023cwx, Malik_2018}.

Traditionally, the inference to constrain the EoS band has been approached within a Bayesian framework, where parametric EoS models are sampled using Markov Chain Monte Carlo (MCMC) techniques to construct posterior distributions consistent with observational data \cite{Steiner_2010, Raaijmakers_2019, Saha:2025don, Steiner_2013, Raithel_2017, Traversi_2020, Thrane_2019}. While robust and statistically well-founded, these methods are computationally expensive and scale poorly with increasing model complexity, often requiring hours to days for convergence. This limitation becomes particularly restrictive in the era of large-scale multi-messenger observations, where rapid inference is essential \cite{Antoniodis_2013, Malik:2022zol, LandryEssick2019GP, Dong2025, PhysRevLett.127.241103}. 
Although Bayesian statistics does a commendable job, it is intrinsically slow and requires substantial computational power. On the other hand, recent advances in machine learning have opened new opportunities for addressing such inverse problems: the direct reconstruction of the underlying microscopic EoS from macroscopic $M$-$R$ observations  \cite{Fujimoto_2018, Fujimoto_2020, Fujimoto_2021, Huang_2024rfg, Armstrong:2025tza, Patra:2026yuv, Patra:2025xtd,carvalho2025neuralposteriorestimationneutron,Soma_2022,Chatterjee:2023ecc,Mukherjee:2026srq}. In particular, conditional invertible neural networks (cINNs) provide a powerful framework for learning bijective mappings between observed data and the corresponding EoS data \cite{ardizzone2019analyzinginverseproblemsinvertible}. Unlike conventional neural networks, cINNs enable exact likelihood evaluation and reversible transformations, making them naturally suited for probabilistic inference tasks. Since observational constraints-such as those from NICER (Neutron star Interior Composition Explorer), do not yield discrete measurements but rather broad, continuous posterior patches in the $M$-$R$ plane, bridging these macroscopic uncertainties to the microscopic EoS requires a model capable of mapping to all probable central values. In this sense, cINNs can be viewed as learning an inferred posterior distribution: once trained, they allow direct sampling of solutions conditioned on observational data, rather than just outputting one single answer. Their underlying design enables them to map all valid possibilities and their likelihoods, providing a fast and comprehensive picture of the results \cite{buchner2021ultranestrobustgeneral, Gandolfi:2019zpj, Somasundaram:2024ykk}.
\par
However, purely data-driven models often lack a guarantee of physical consistency, which is critical in astrophysical applications. To address this, we incorporate physics-informed regularisation (PIR) into our framework. PIR embeds physical laws directly into the training process by penalizing violations of governing constraints \cite{RAISSI2019686, karniadakis2021physics}. In our implementation, we enforce causality and thermodynamic stability by constraining the squared speed of sound, $c_s^2 = \frac{dp}{d\epsilon}$, to lie within physical bounds $(0 \leq c_s^2 \leq 1)$ \cite{PhysRevLett.114.031103, 2020NatPh..16..907A, Brandes:2022nxa, Altiparmak:2022bke, Roy:2022nwy, ecker_2022, Kurkela_2014, Chatterjee:2023ecc}.

In this work, we develop a hybrid cINN+PIR architecture that maps $M$-$R$ posterior contours to the most probable regions in central energy density-central pressure space. By training on physically consistent data and enforcing constraints through the PIR component, our model learns a structured, interpretable representation of the inverse mapping. To validate our architecture against current astrophysical constraints, we apply our model to a speed-of-sound interpolated (SoSI) EoS dataset. We demonstrate the network's ability to seamlessly integrate and satisfy joint observational contours, specifically mapping 95\% confidence macroscopic constraints of the gravitational wave event GW170817 alongside the 95\% confidence $M$-$R$ posterior contours derived from NICER observations of PSR J0030+0451, J0437-4715, J0614-3329, and J0740+6620 \cite{PhysRevLett.121.161101,Miller_2019,Choudhury_2024,Mauviard_2025,Miller_2021, GW170817, Nicholl:2021rcr, Riley_2021, Demorest_2010}.

The structure of this paper is as follows. In Section \ref{sec:formalism} (Formalism), we describe the architecture of the cINN with the incorporation of PIR and the EoS data generation. In Section \ref{sosi} (Results and Discussion), we present results of synthetic and real observational data in the model's PIR credibility by generating causal and thermodynamically stable outputs in $p$-$\epsilon$ space. In the same section, we further test the model's ability to retain the physically valid EoS parameter space, followed by a predictive scan for optimal future targets. Finally in Section \ref{sec:summary} (Summary and Conclusion), we summarise our work and draw out the important conclusions. 

\section{Formalism: Model Architecture and Data Generation}
\label{sec:formalism}

Conditional Invertible Neural Networks (cINNs) are a class of deep generative models based on normalising flows. By definition, normalising flows are mathematical frameworks designed to learn complex, high-dimensional probability distributions by transforming a simple, tractable base distribution (such as a standard multivariate Gaussian) through a sequence of bijective, differentiable functions \cite{papamakarios2021normalizingflowsprobabilisticmodeling, rezende2016variationalinferencenormalizingflows}. In the context of compact object astrophysics, cINNs provide a mathematically rigorous framework for solving highly degenerate inverse problems. Rather than predicting a single deterministic point estimate, the network learns a conditional, invertible mapping between a latent Gaussian space $z$ (as described from Eq. \ref{forwardeq}) and the physical target parameters $x$ (in this work, the central energy density ($\epsilon_c$), central pressure ($p_c$), and speed of sound at the center of the star ($c_{s,c}^2$), conditioned on macroscopic observables $c$-the $M$-$R$ posterior contours. This probabilistic approach is exceptionally well-suited for multi-messenger astronomy. Observational constraints from X-ray profiling (e.g. NICER) and gravitational wave detectors do not yield exact $M$-$R$ coordinates; rather, they provide extended 2-D posterior probability contours \cite{Abbott_2019, Riley_2019}. By conditioning the flow on these macroscopic $M$-$R$ confidence regions, the cINN directly generates the corresponding conditional posterior distribution $p(x \mid c)$ for the stellar central parameters. The step-by-step construction and optimisation of this architecture are detailed below, while the model design and the data flow pipeline are illustrated in Figure \ref{fig:cinn_model_pipeline} in Appendix \ref{app:model_pipeline}.

\subsection{Data Representation and Contour Encoding}\label{data_rep}
During training, the conditioning information ${c}$ is constructed as a 200-dimensional feature vector from flattened $M$-$R$ values, each with 100 discrete $M$-$R$ coordinates. The feature vector is defined in the real coordinate space ($\mathbb{R}$) as $c \in \mathbb{R}^{200}$. Extracted from the physically valid, TOV-solved equations of state \cite{PhysRev.55.364} is the physical target vector $x$ representing the central EoS values of stars, defined as 
    
     \begin{equation}\label{x}
         x = (\log \epsilon_c, \log p_c, c_{s,c}^2) \in \mathbb{R}^3
     \end{equation}

    To ensure stable convergence, both inputs $c$ and targets $x$ undergo global Z-score normalization-a process which standardizes the data by subtracting the mean of the dataset and dividing by the global standard deviation for each dimension, yielding zero-mean and unit-variance features, $c'$ and $x'$. Furthermore, to prevent the normalizing flow \cite{Kobyzev_2021} from memorizing discrete training manifolds-which causes artificial likelihood spikes, a Gaussian noise $\epsilon \sim \mathcal{N}(0, \sigma_{noise}^2)$ is injected into the normalized targets during the forward pass. Here, $\mathcal{N}$ denotes a normal distribution with a mean of zero, and the variance $\sigma_{noise}^2$ is parameterized using the previously computed standard deviation of the target variables. Rather than conditioning the flow directly on the flattened 200-dimensional vector, a 1-D Convolutional Neural Network (CNN) is employed as a geometric feature extractor \cite{lecun2015deep}. Operating on two standardized channels ($M$ and $R$), the convolutional layers act as local spatial filters, automatically detecting physically relevant macroscopic geometries along the contour, such as local slope, curvature, boundary width, and shape anisotropy. The network compresses this spatial information into a dense, 32-dimensional context embedding $c_{emb} \in \mathbb{R}^{32}$. A 32-dimensional space acts as an optimal balanced filter-preventing the memorization of discrete coordinate noise while retaining sufficient capacity to map complex contour geometries to the 3-dimensional physical target space (Eq. \ref{x}). This embedding serves as the continuous conditioning vector for the invertible blocks (detailed in the next section), providing a highly optimized summary of the observational constraints.
    
     \subsection{Invertible Architecture and Likelihood Optimization}
     The core of the cINN is a sequence of bijective transformations designed to map the complex, unknown distribution of the normalized physical targets $x'$ into a simplified, tractable probability space. Specifically, conditioned on the context embedding $c_{emb}$, we map $x'$ to a continuous latent variable $z$ that is strictly constrained to follow a standard multivariate Gaussian distribution, $z \sim \mathcal{N}(0, I)$. We guarantee exact invertibility and a computationally tractable Jacobian determinant by utilizing a sequence of eight affine coupling layers \cite{dinh2017densityestimationusingreal}.
    
     In each coupling layer, the input vector is partitioned into two disjoint sub-vectors, $x_a$ and $x_b$. The unmasked partition $x_a$ is concatenated with the context embedding to form $u = [x_a, c_{emb}]$. This combined vector passes through an internal sub-network of sequential linear transformations and ReLU activations to output a raw vector $v$. To prevent exploding gradients during optimization, this output is bounded via a scaled hyperbolic tangent:
    
    \begin{equation}
        (s_\theta, t_\theta) = 2 \cdot \tanh(v)
    \end{equation}
    
    where $\theta$ denotes the trainable weights and biases of this internal sub-network. These split vectors define the scale ($s_\theta$) and translation ($t_\theta$) for the affine transformation. Within this affine transformation, the scaling term $s_\theta$ controls the variance by stretching or compressing the parameter space, while the translation term $t_\theta$ dictates the spatial shift of the distribution's mean. Scaling and shifting allow the model to map complex physical distributions into tractable Gaussians while guaranteeing exact mathematical invertibility and computationally cheap Jacobian evaluations for efficient training. These are applied to the masked partition $x_b$, while $x_a$ propagates forward unchanged:
    
    \begin{equation}
    \begin{aligned}
        &x_b' = x_b \odot \exp(s_\theta) + t_\theta \\
        &x_a' = x_a
    \end{aligned}
    \end{equation}
    
    where $\odot$ denotes element-wise multiplication. Because $s_\theta$ and $t_\theta$ depend solely on the untransformed $x_a$, the inverse operation requires only simple subtraction and division, without ever needing to invert the sub-network $\theta$ itself. Coordinate splitting is permuted between successive blocks to ensure all dimensions interact over the depth of the network. The total transformation is the exact mathematical composition of these sequential mappings. Denoting the operation of the $i$-th coupling layer as $f_i$, the complete forward pass is formulated as:
    
    \begin{equation} \label{forwardeq}
    \begin{split}
        z &= f_\theta(x' \mid c_{emb}) \\
          &= f_8 \Big( f_7 \big( \dots f_1(x' \mid c_{emb}) \dots \mid c_{emb} \big) \mid c_{emb} \Big)
    \end{split}
    \end{equation}
    
    This strictly bijective mapping allows for the exact conditional probability density to be evaluated via the change of variables formula. The network is optimized by minimizing the negative log-likelihood (NLL) of the training data:
    
    \begin{equation} \label{nll_loss}
        \mathcal{L}_{NLL} = \frac{1}{2} ||z||^2 - \log |\det J|
    \end{equation}
    
    where $J_{ij} = \frac{\partial z_i}{\partial x'_j}$ is the Jacobian matrix of the full transformation. This objective function balances two competing geometric effects: the $||z||$ latent prior constraint pulls the mapped physical parameters toward the high-probability center of the Gaussian latent space, while the log-Jacobian volumetric regularizer explicitly accounts for density changes, penalizing infinite phase-space compressions to prevent mode collapse.

    \subsection{Physics-Informed Regularization (PIR Framework)}

    While the NLL objective ensures the network learns the empirical data distribution, purely data-driven generative models offer no intrinsic guarantees of physical consistency. Without constraints, the continuous density learned by the flow might extrapolate unphysical states outside the immediate training manifold. To enforce adherence to the fundamental principles of relativistic stellar structure, we embed a PIR constraint directly into the training loop. Specifically, the generated EoS must satisfy two strict macroscopic bounds governing the squared speed of sound, $c_{s}^2 = \frac{dp}{d\epsilon}$ \cite{Alford_2013, Han_2019}:
    
    \begin{itemize}
        \item \textbf{Causality:} The speed of sound within the dense medium must not exceed the speed of light in vacuum. In natural units, this requires $c_{s}^2 < 1$. 
        \item \textbf{Thermodynamic Stability:} The matter must be dynamically stable against microscopic collapse, requiring a positive bulk modulus, hence $c_{s}^2 \geq 0$.
    \end{itemize}
    
    To actively penalize violations of these laws during training, we leverage the exact reversibility of the cINN architecture. In each iteration, we sample random latent vectors, denoted as $z_{sample}$, from the target prior distribution. This prior is defined as a standard multivariate Gaussian $\mathcal{N}(0, I)$, representing a space with zero mean and an identity covariance matrix. We process these samples through the inverse network mapping, conditioned on the context embedding:
    
    \begin{equation}
        x'_{gen} = f_\theta^{-1}(z_{sample} \mid c_{emb})
    \end{equation}
    
    This inverse pass maps the abstract Gaussian vectors back into the network's predicted physical parameters, yielding the generated output $x'_{gen}$ by inverting the forward pass from equation \ref{forwardeq}. However, $x'_{gen}$ exists in the normalized, zero-mean data space used for training. Because the fundamental physical limits of $0$ and $1$ apply exclusively to the true, unscaled values of the speed of sound, we must reverse the Z-score normalization process (as described in section \ref{data_rep}) before the constraints can be evaluated. We denormalize the generated output to recover the real physical parameters ($x_{real}$) by reapplying the training dataset's statistical standard deviation ($\sigma_x$) and mean ($\mu_x$):
    
    \begin{equation}
        x_{real} = x'_{gen} \odot \sigma_x + \mu_x
    \end{equation}
    
    Extracting the predicted $c_{s,c}^2$ from the unscaled vector $x_{real}$, we formulate the physics-informed loss ($\mathcal{L}_{phy}$) using a piecewise penalty function. This function strictly activates only when predictions fall outside the physically permissible bounds:
    
    \begin{equation}\label{phy_loss}
        \mathcal{L}_{phy} = \max(0, c_{s,c}^2 - 1) + \max(0, -c_{s,c}^2)
    \end{equation}
    
    This formulation serves as a highly effective physical filter. If the network generates valid EoS parameters, $\mathcal{L}_{phy}$ evaluates to exactly zero. However, if the network attempts to map the latent space to unphysical, acausal, or unstable regions, the resulting gradients heavily penalize the network weights. 

    \subsection{Total Loss and Optimization}
    
    The network is trained end-to-end by minimizing a joint objective function that balances the exact data likelihood with the physical constraints. By combining Eq. \ref{phy_loss} with the negative log likelihood from Eq. \ref{nll_loss}, the resulting cINN+PIR framework gives an inherent posterior distribution that is both data-driven and strictly bounded by fundamental physics. This combination gives a total loss equation
    
    \begin{equation}
        \mathcal{L}_{total} = \mathcal{L}_{NLL} + \lambda \mathcal{L}_{phy}
    \end{equation}
    
    where $\lambda$ is a penalty weighting hyperparameter that heavily penalizes acausal or unstable generative passes. The penalty weight $\lambda = 10.0$ was empirically chosen because it is large enough to strictly enforce physical boundaries without overpowering the network's ability to learn the data distribution. To optimize the network, the parameters $\theta$ of the internal sub-networks across all affine coupling layers are updated via backpropagation. Backpropagation is the algorithmic process neural networks use to learn from their mistakes by working backward from the output to calculate exactly how much each internal weight must be adjusted to minimize future errors. The optimization process builds upon the fundamental gradient descent step \cite{kingma2017adammethodstochasticoptimization}:

    \begin{equation}
         \theta - \eta \nabla_\theta \mathcal{L}_{total} \rightarrow \theta
    \end{equation}

\begin{figure}[ht!]
    \centering
    \includegraphics[width=\linewidth]{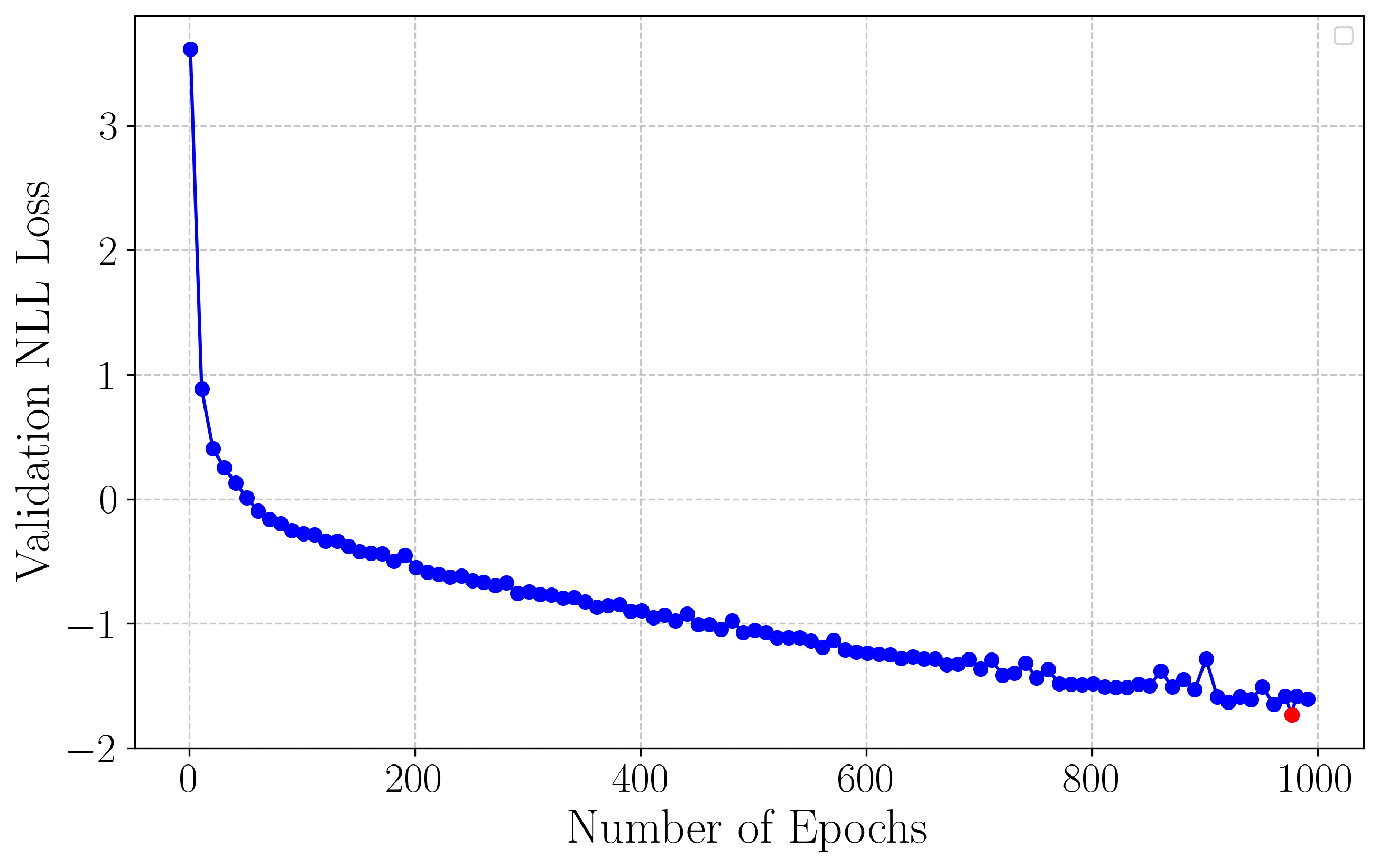}
    \caption{Evolution of the validation NLL loss ($\mathcal{L}_{total}$) over successive epochs. The transition to a negative loss signifies the network's convergence onto a highly confident, localized probability density mapping.}
    \label{fig:loss_model}
    \end{figure}

    Conceptually, $\nabla_\theta \mathcal{L}_{total}$ represents the gradient-a mathematical vector indicating the direction of steepest increase in the model's error. By subtracting this gradient from the current network parameters ($\theta$), the model takes a calculated step in the exact opposite direction to iteratively minimize the loss. The learning rate, $\eta$, acts as a step-size multiplier controlling how aggressively the parameters adjust during this update. In our model, the base learning rate is set to $\eta = 10^{-4}$. This value of $\eta$ was established through experimental testing, providing the most stable convergence and optimal loss minimisation. Because the cINN outputs a continuous probability density function (PDF) rather than a discrete probability mass, density values are not strictly bounded by $1$. Much like compressing a fixed volume of fluid into a narrower column increases its height, as the network becomes more confident, it concentrates the total probability volume (which must strictly integrate to unity) into a progressively smaller parameter space. This causes the localised density values to exceed $1$ by a significant amount. Mathematically, the logarithm of a density greater than $1$ yields a positive value; consequently, the NLL objective becomes negative. As illustrated in Figure \ref{fig:loss_model}, the loss initially takes a positive value, indicating an uncalibrated network with broad, highly uncertain mappings. As the network learns the underlying geometric and physical correlations, the predicted conditional posterior distribution, denoted as $p(x\mid c_{emb})$, systematically narrows. The probability density sharply peaks around the true stellar interior parameters, driving the NLL aggressively into the negative regime. Therefore, a highly negative NLL is a desirable outcome, serving as a direct mathematical proxy for extreme model confidence and precision.

   To prevent overfitting to the training manifold, we implement an early-stopping protocol evaluated on an independent validation set. Training is halted, and the optimal network weights are preserved, when the validation NLL reaches its saturated minimum at $\mathcal{L}_{NLL} = -1.73$, ensuring generalization to unseen astrophysical data.

    \subsection{Data Generation}
    \label{sec:data}

    For the training data, we use a total of 90,000 EoSs comprising of the two types Monotonic and Non-Monotonic \cite{Verma:2025dez} each type generated in statistical bulk using the SoSI method \cite{Alford_2005,Kojo_2016,Baym_2019,Annala_2020, Nath:2023gmu}. The SoSI method constructs these EoSs by modeling the squared speed of sound as a function of energy density. Specifically, each EoS is generated using a baseline crust parameter ($\Gamma_{CET}$) \cite{1971ApJ...170..299B,Hebeler_2013, Kurkela2010pQCD, Fraga:2013qra, Tews_2013, Lattimer_2016} and four discrete speed-of-sound anchor points ($C_1$, $C_2$, $C_3$, $C_4$) \cite{Annala_2020,Verma:2025vkk,Tews_2018}. For the monotonic EoS class, $c_{s}^2$ strictly increases up to the neutron star core density. In contrast, for the non-monotonic class, $c_{s}^2$ reaches a peak and experiences a subsequent dip before reaching core density, allowing the dataset to comprehensively simulate diverse, physically plausible behaviors within the stellar interior. The built EoSs are solved using TOV equations \cite{PhysRev.55.364,PhysRev.55.374} to generate $M$-$R$ sequences and each star's corresponding $\epsilon_c$, $p_c$ and $c_{s,c}^2$ is saved. These central value relations with respective $M$-$R$ values serve as input during the training of our cINN model while $c_{s,c}^2$ is used as a physical check in the PIR part of our model architecture that directly influence the learning of cINN. 
    
\section{Results and Discussion}
\label{sosi}

Having established the theoretical framework and convergence stability of the hybrid cINN+PIR architecture, we now evaluate its predictive performance. To ensure the reliability of our predictions and reduce the structural uncertainty inherent to complex neural networks, we employ a deep ensemble sampling methodology \cite{lakshminarayanan2017simplescalablepredictiveuncertainty} - a technique where multiple independent neural networks are trained to solve the same problem so their predictions can be aggregated. Because the optimization landscape of normalizing flows is highly non-convex, a single model's convergence can be sensitive to stochastic, or inherently random, variations during training. To counteract this, we independently initialize and train an ensemble of five identical cINN+PIR models. For each individual model, we assign a different random seed. By varying this seed, we guarantee that each of the five models receives a uniquely randomized starting set of network weights ($\theta$) and experiences a distinct shuffling of the training data batches. Upon reaching convergence criteria via our validation early-stopping protocol, the five models are frozen. To perform inference on a given $M$-$R$ contour, the input condition $c_{emb}$ is passed through all five independent networks simultaneously. The generated target variables are then aggregated to form a joint conditional posterior distribution. This ensemble approach averages out initialization artifacts and single-model biases, much like deploying multiple independent random-walk chains in Bayesian inference \cite{Foreman_Mackey_2013}. Therefore, the maximum probability zones identified below reflect robust global optima in the EoS parameter space.

\subsection{Synthetic contour inference and physical validation of the cINN posterior}
\label{testinGT}
\begin{figure}[ht!]
    \centering
        \includegraphics[width=\linewidth]{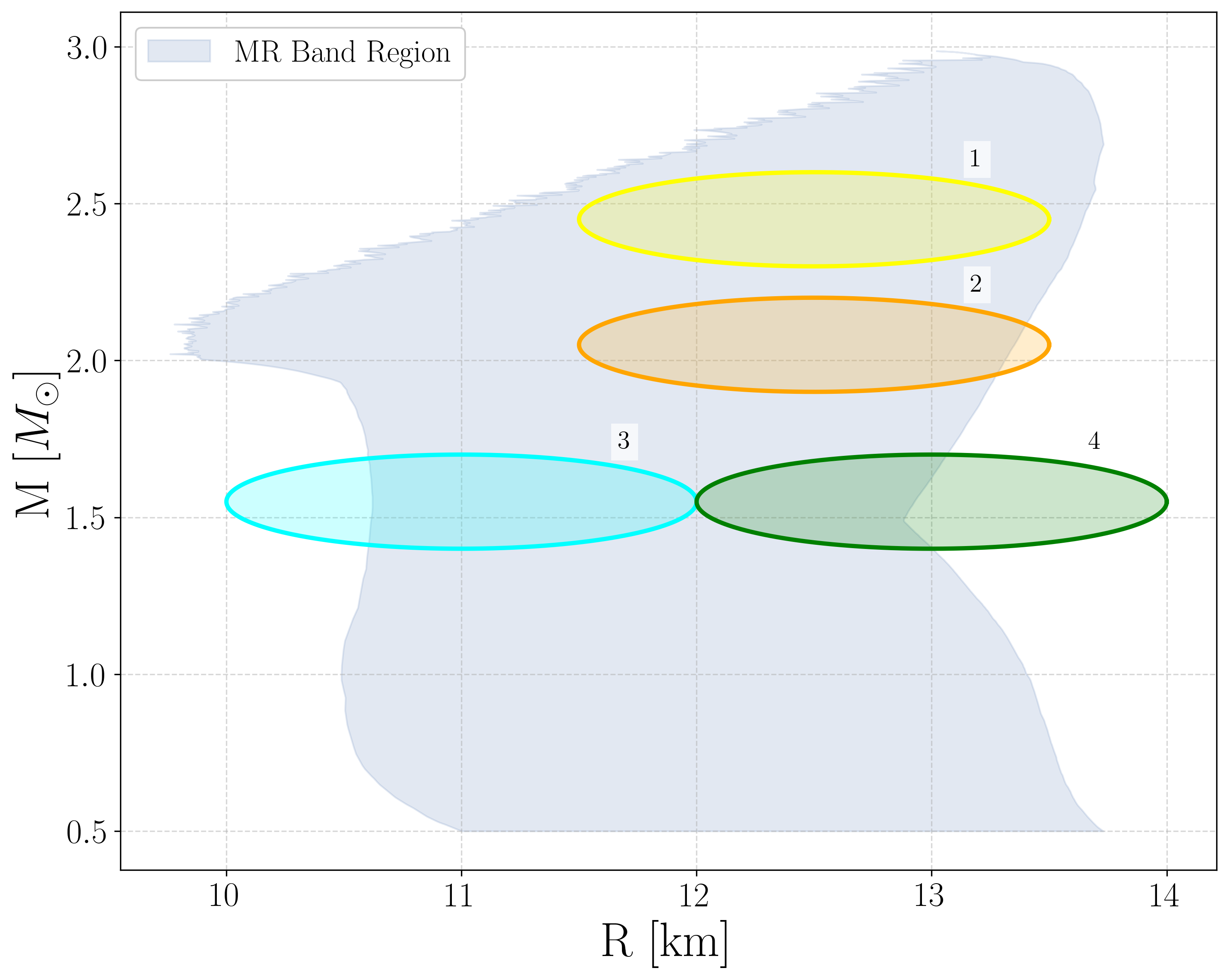} 
        \includegraphics[width=\linewidth]{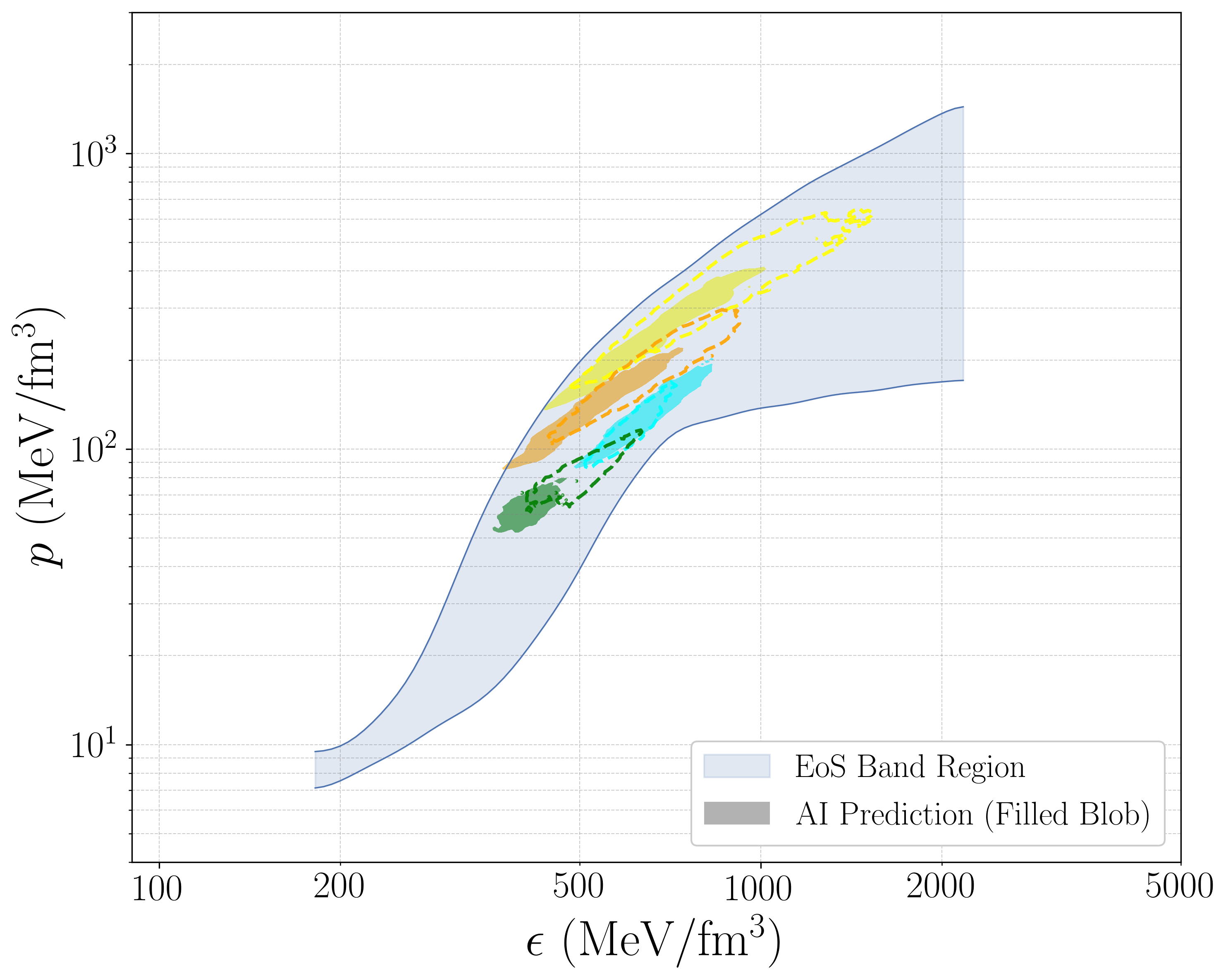}
    
    \caption{\textit{Upper panel :-} Synthetic stars utilized as NICER mock observations overlaid on $M$-$R$ band region of the corresponding EoSs used in neural network training. \textit{Lower panel :-} Most probable central region posteriors corresponding to each mass-radius contour shown on a logarithmic scale. Filled region represents the cINN output while dashed contours represent the parameter space of all known theoretical $p_c$-$\epsilon_c$ values consistent with the TOV solved $M$-$R$ data corresponding to the synthetic contour regions in Fig. \ref{fig:synthetic_contour} (upper panel). Background color band indicates the range of EoS models (monotonic and non-monotonic) utilized during neural network training.}
    \label{fig:synthetic_contour}
    
\end{figure}

To establish an understanding of the model's mapping capabilities, we first evaluate the cINN on four synthetic $M$-$R$ contours designed to simulate the structural characteristics of NICER observations where the contours span a fixed elliptical dimension of mass $\Delta M = 0.3 M_\odot$ and radius $\Delta R = 2$ km as shown in Figure \ref{fig:synthetic_contour} (upper panel). The resultant central EoS contours, specifically, the median (50th percentile) of the cINN predictive distributions, which best aligns with our comparative baseline, are presented in Figure \ref{fig:synthetic_contour} (lower panel). This test demonstrates the network's ability to transition directly from the observable $M$-$R$ space to the high-energy central EoS space $(\epsilon_c, p_c)$. Crucially, this approach performs the inversion using machine learning, circumventing the need for the explicit physical construction of each type of EoS and computationally expensive integration of the TOV equations to solve them.

As shown in Figure \ref{fig:synthetic_contour}, the cINN captures the most probable posterior regions consistent with the physical relationship between a star's macroscopic properties and its internal structure. The high-mass configurations (stars 1 and 2) yield cINN posteriors localized at appropriately high central energy densities and pressures, a defining characteristic of massive compact objects. Moreover, the model has also learned to constrain the vast regions of EoS of the high mass stars (especially stars 1 and 2) and found the most probable zones of the central EoS values to be significantly smaller than the posterior regions (dashed lines) of training EoS data corresponding to the TOV solved $M$-$R$ curve values that are situated inside the $M$-$R$ contour regions in Figure \ref{fig:synthetic_contour} (upper panel). In particular, star 4 (green contour) exhibits interesting behaviour: being an intermediate-mass star with a large radius, this star must fundamentally possess lower EoS parameters due to its low compactness. Moreover, a major part of this star contour falls outside the known data region of $M$-$R$, which was used in training (Figure \ref{fig:synthetic_contour} upper panel). This contour, therefore, tests the model's ability to extrapolate beyond the regions where it has never seen data during training. As a result, the cINN posterior for this unknown region extends beyond the training area contour curves plotted in dashed (Figure \ref{fig:synthetic_contour}, lower panel) in the lower region of EoS space, reflecting the behaviour of stars with lower compactness.

M-R curves reflect the properties of EoS. Its slope \(dM/dR\) is known to reflect the stiffness properties of its EoS. Based on its slope an M-R curve can be front-bended (\(dM/dR <0\)), as well as back-bended (\(dM/dR \nless 0\)) \cite{Tan:2021nat, Ferreira:2024hxc,Bauswein:2025dfg}. Configurations 1,2 and 4 are populated with such back-bended MR curves. These curves are associated with EoS that exhibit early stiffening, which results in smaller values of central density. Conversely, for front-bended MR curves, the stiffening of EoS occurs late, at higher density, which results in higher values of central densities \cite{Mukherjee:2026srq}. Our cINN cleanly reproduces this behavior (Figure \ref{fig:synthetic_contour}, lower panel). In particular, configuration 4 is correctly mapped to a lower-density region, whereas configurations 1 and 2 exhibit an extended mapping, reflecting the presence of both front- and back-bended MR curves.

\begin{figure}[h]
    \centering
    \includegraphics[width = \linewidth]{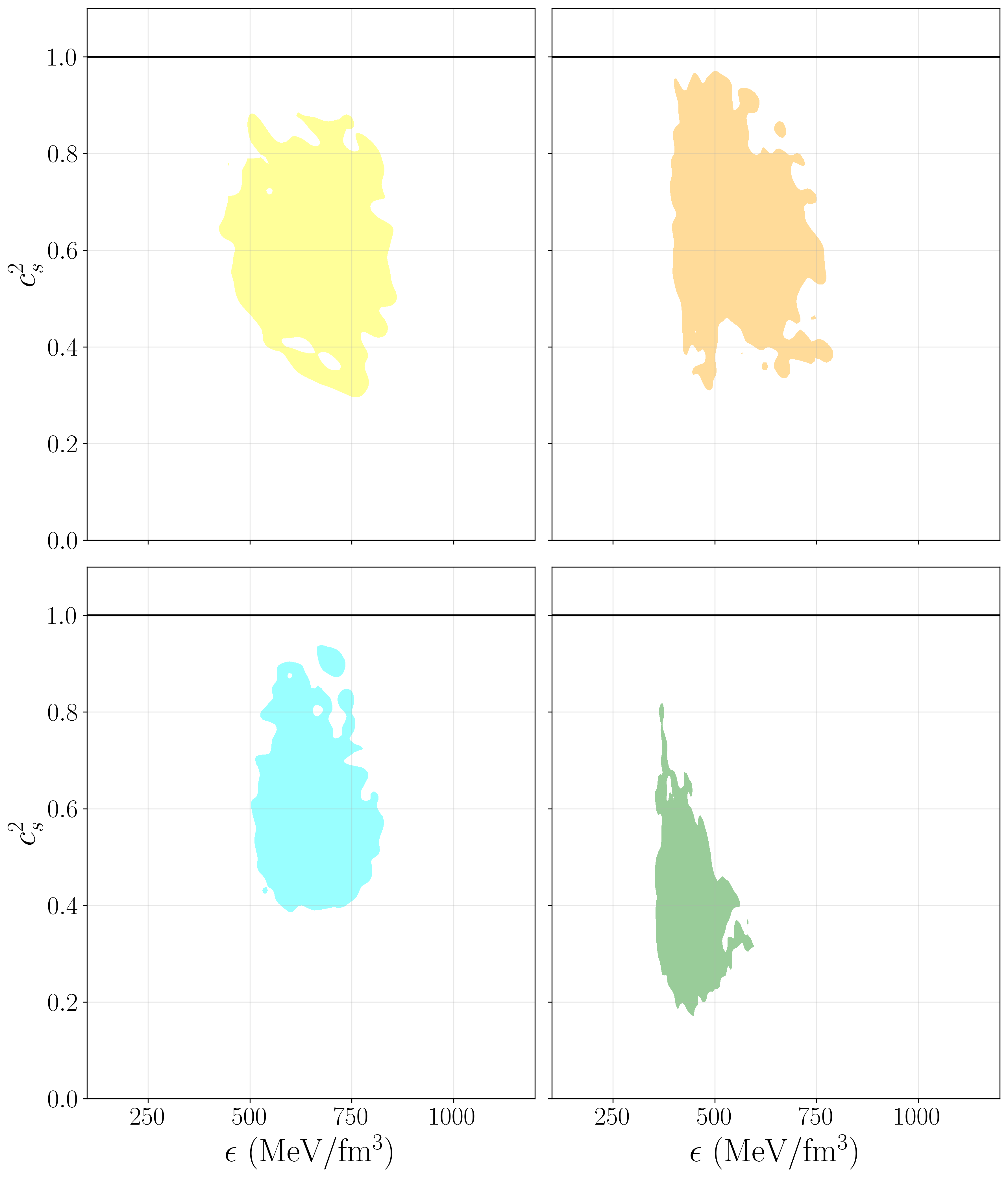}
    \caption{Posterior distribution of $c_{s,c}^2$ ($c_s^2$ at the center of the star) as a function of central energy density ($\epsilon_c$) generated by the cINN+PIR ensemble. Each colour zone corresponds to the star in Fig. \ref{fig:synthetic_contour} (upper panel). The physical constraints explicitly prohibit predictions from exceeding the causal limit ($c_{s,c}^2 = 1$) described by the black horizontal line.}
    \label{fig:cs2_vs_E}
\end{figure} 

Conventionally, the sound speed is obtained by evaluating the local derivative of a continuous equation of state (EoS), $c_s^2=\partial p/\partial\epsilon$. This poses a fundamental challenge for generative models such as cINNs, which do not produce continuous EoS curves. Instead, they learn a conditional probability distribution and generate an ensemble of independent point-wise predictions in the thermodynamic phase space, rendering post-inference numerical differentiation impossible.

To overcome this limitation, we reformulate the role of the sound speed within the network. Rather than treating $c_{s,c}^2$ as a derived quantity, we promote it to a primary thermodynamic variable that is co-predicted together with the central pressure, $p_c$, and central energy density, $\epsilon_c$. During data generation, the training EoSs are continuous, allowing $c_{s,c}^2$ to be computed explicitly at every thermodynamic state. The cINN therefore, learns the joint conditional distribution $[\mathbb{P}(\epsilon_c,p_c,c_{s,c}^2,|,c_{M-R})]$,
where $c_{M-R}$ denotes the observed mass--radius contour.

This formulation enables the direct application of Physics-Informed Regularisation (PIR). During training, custom penalty terms enforce the physical bounds $0\le c_s^2\le1$, corresponding to thermodynamic stability and causality. Since $\epsilon_c$, $p_c$, and $c_{s,c}^2$ are learned jointly within a shared latent representation, these penalties constrain the entire inverse mapping, ensuring that the inferred central pressure and energy-density posteriors are sampled only from physically admissible regions. Consequently, although the generated samples are independent points rather than continuous curves, each point carries its correct local thermodynamic information inherited from the training EoSs. As a result, the cINN+PIR reconstructs physically consistent posterior distributions in the $(\epsilon_c,c_s^2)$ plane while keeping the entire generated ensemble strictly within the causal and thermodynamically stable region, as illustrated in Fig.~\ref{fig:cs2_vs_E}. 

The distribution of $c_{s}^2$ for the four test configurations reveals that cINN was able to reproduce the EoS properties one would expect for both more compact and massive stellar models. Comparing the configurations of 3 and 4, lower radii associated with configurations of 3 allow them to be more compact than 4, which requires greater stiffness, as has been reproduced by significantly higher $c_{s}^2$ values shown in the lower panel of Figure \ref{fig:cs2_vs_E}. Also, comparing the configurations of 1 and 2, the distribution of $c_{s}^2$ for the massive models (configuration 1) is indeed lower than that of configuration 2. This is again a known physical feature \citep{ecker_2022, Saha:2024swd} as the massive neutron stars are expected to undergo phase transitions (smooth), resulting in softer cores indicated by relatively smaller values of $c_{s}^2$ compared to configuration 2, shown in the upper panel of Figure \ref{fig:cs2_vs_E}.

\subsection{Multi-observation consistency and theoretical baseline recovery}

\begin{figure}[h]
    % \centering
    \includegraphics[width=\linewidth]{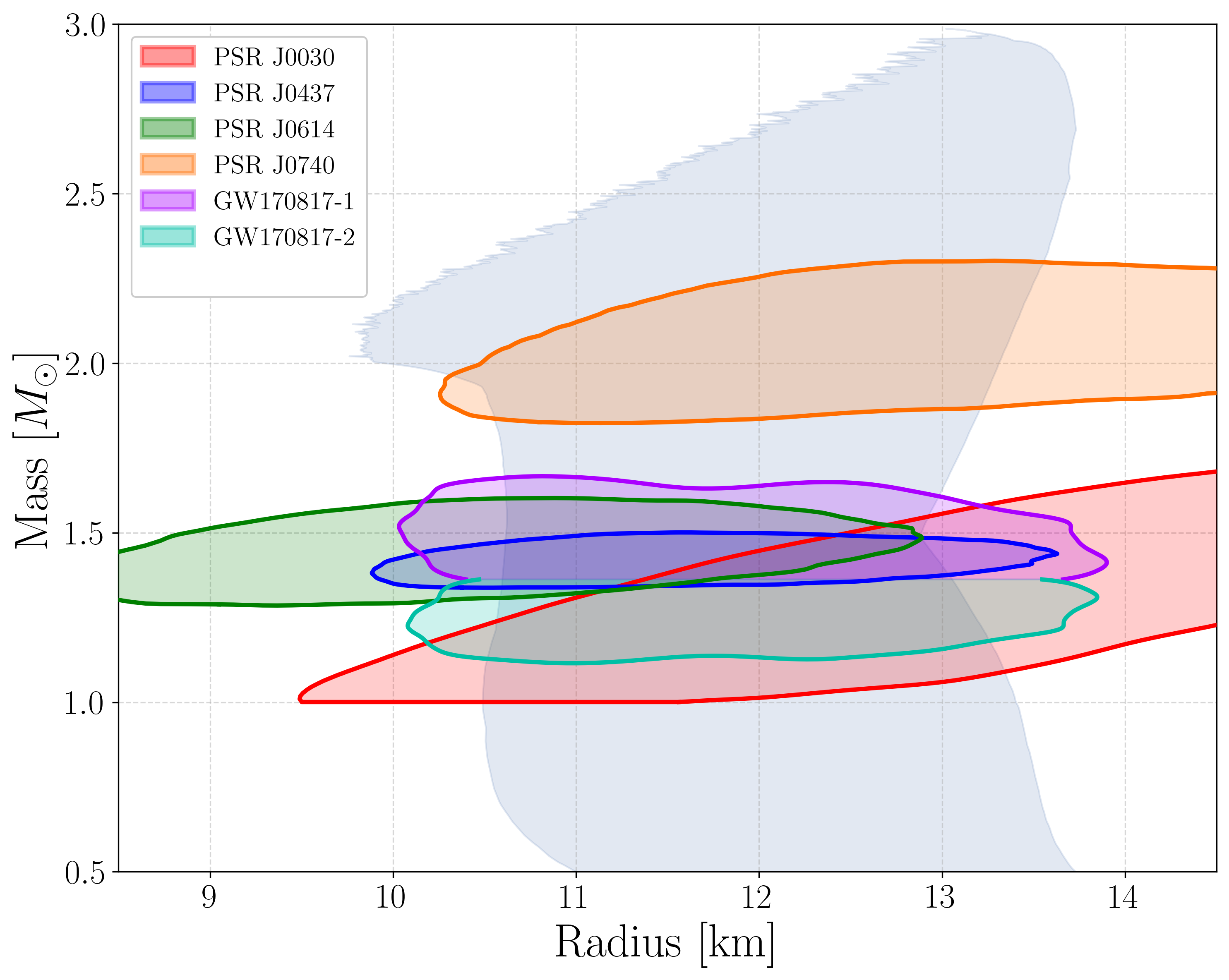}
    \caption{Observational $M$-$R$ contours derived from NICER measurements and the GW170817 binary neutron star merger.}
    \label{fig:Nicer_contours}
\end{figure}

Following validation on synthetic contours (Section~\ref{testinGT}) and demonstrating that the PIR guarantees causal and thermodynamically stable predictions, we apply the cINN+PIR framework to observational data. Specifically, we use the mass--radius constraints from NICER (PSR J0030+0451, J0437$-$4715, J0614$-$3329, and J0740+6620) together with the component-star constraints from the binary neutron-star merger GW170817, shown in Fig.~\ref{fig:Nicer_contours}.

\begin{figure*}[ht]
    \centering
    \includegraphics[width=\linewidth]{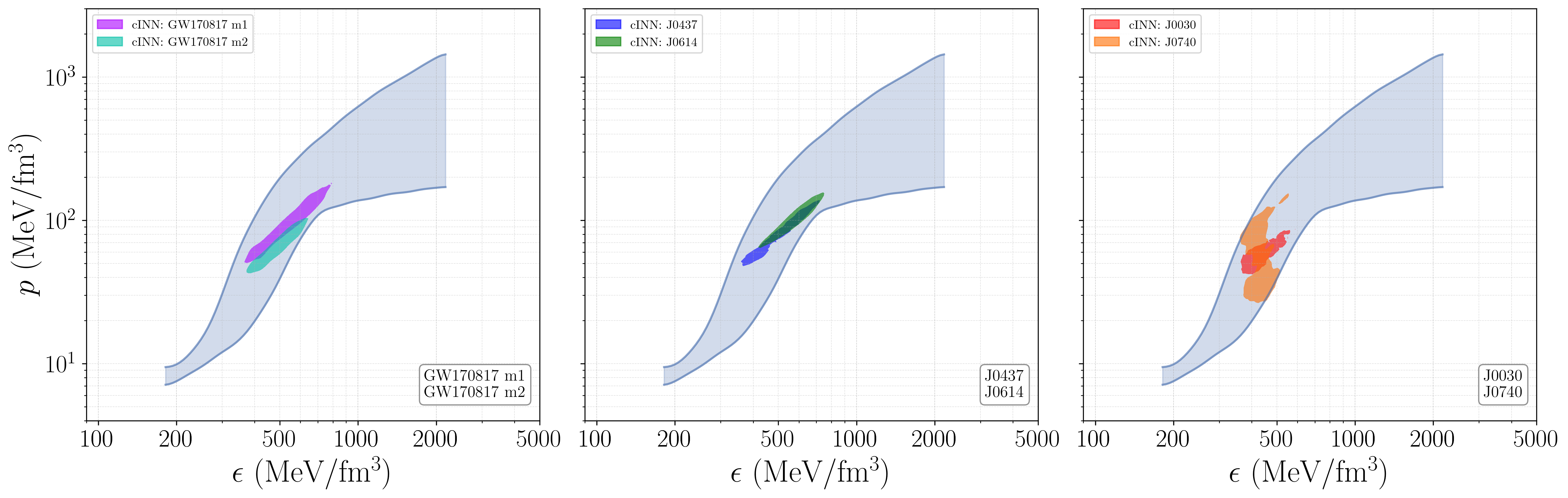}
    \caption{Individual cINN $p_c$-$\epsilon_c$ posterior regions inferred from each NICER and gravitational wave observations corresponding to Fig. \ref{fig:Nicer_contours}. On a logarithmic scale, each panel shows the 50th-percentile KDE boundary of the five-member ensemble prediction conditioned on the respective observation.}
    \label{fig:cinn_blobs}
\end{figure*}

To assess the ability of the cINN+PIR to recover the physically allowed EoS space, we compare its predictions with the baseline speed-of-sound interpolation (SoSI) ensemble satisfying all six observational constraints simultaneously. Applying the cINN independently to each observation produces six posterior regions in the $(\log_{10}\epsilon_c,\log_{10}p_c)$ plane (Fig.~\ref{fig:cinn_blobs}). An EoS from the training catalogue is retained in the final cINN-filtered ensemble only if its $(\epsilon_c,p_c)$ trajectory intersects all six posterior regions simultaneously, providing a stringent multi-observation consistency criterion.

The resulting cINN-filtered envelope retains $84.2\%$ of the baseline SoSI area ($0.6683$ versus $0.7941$ dex$^2$), with a mean width retention of $84.9\%$ and an Intersection over Union (IoU) of $84.4\%$. The close agreement demonstrates that the cINN faithfully reconstructs the physically allowed EoS region while introducing a nearly uniform reduction in uncertainty across the entire density range, rather than preferentially constraining a particular density regime. Throughout this work, logarithmic distances and areas are expressed in dex, where 1 dex corresponds to one order of magnitude.

The mapped posterior regions in the $(p,\epsilon)$ plane represent the range of central pressures and energy densities compatible with each individual observation. Their morphology directly reflects the structure of the corresponding mass--radius contours. The substantial overlap between the GW170817 primary component, PSR J0437$-$4715, and PSR J0614$-$3329 arises from the overlap of their corresponding regions in the mass--radius plane. In contrast, the GW170817 secondary component occupies the lower boundary of the $(p,\epsilon)$ envelope, consistent with its lower compactness. The posterior for PSR J0740+6620 exhibits a pronounced vertical elongation. Although it is reconstructed from the same mixture of front-bended and back-bended mass--radius solutions, its comparatively high mass and broader radius range correspond to a wider range of compactness values. Reproducing these observations, therefore, requires central energy densities spanning EoSs with significantly different stiffness, giving rise to the extended vertical structure in the $(p,\epsilon)$ plane.

\subsection{Predictive Constraints: Identifying the Optimal Targets for Future Observation}
\label{sec:heatmap_3R_01M}

\begin{figure*}[ht]
    \includegraphics[width = \linewidth, height = 12 cm]{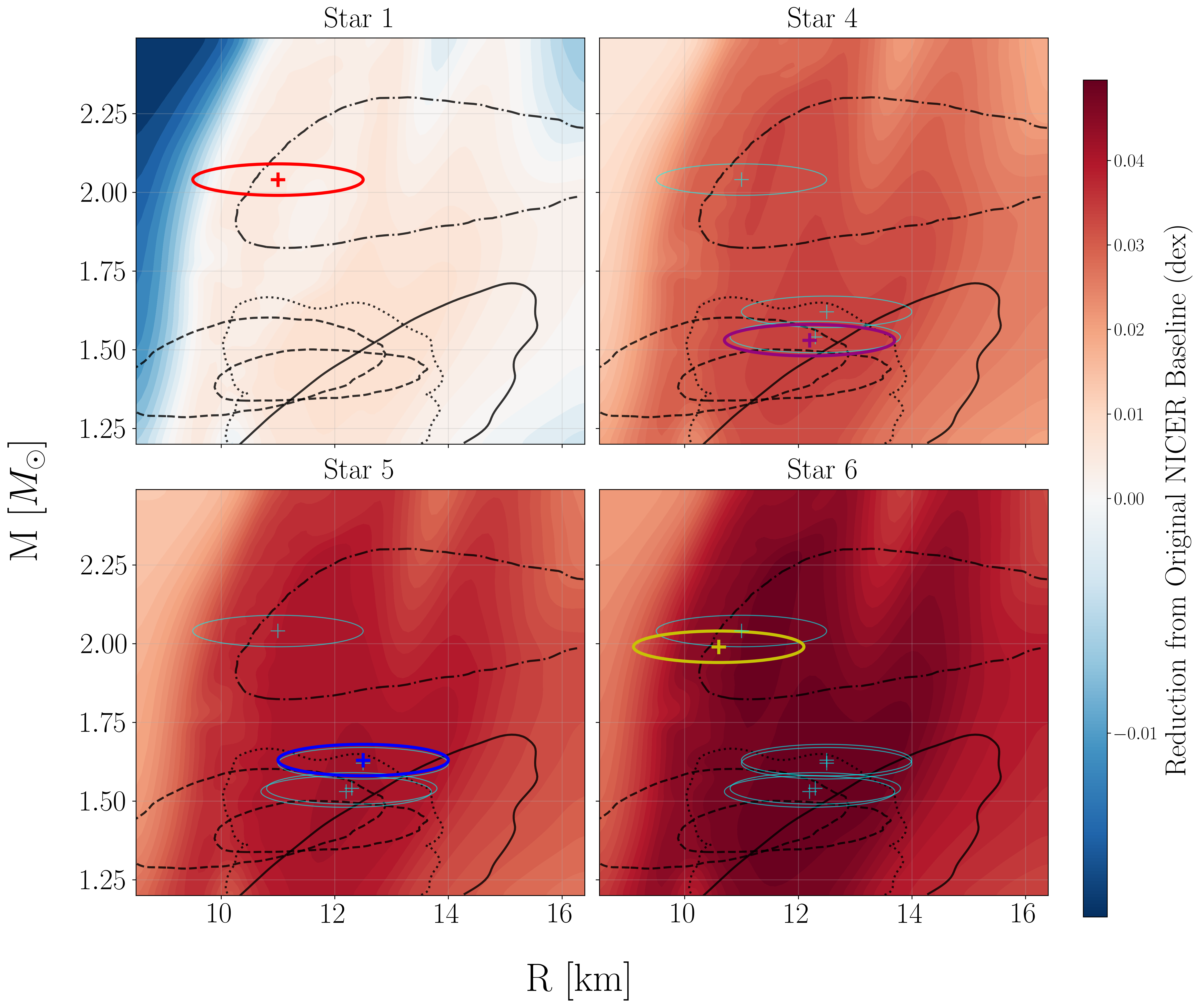}
    \caption{NICER baseline EoS band shrinkage across the mass--radius ($M$--$R$) plane during the sequential search for six optimal observational targets. The heatmaps show the reduction in the EoS band width (in dex) obtained by placing a hypothetical mass--radius measurement at each grid point. As the sequential search progresses, increasingly darker regions indicate a larger reduction relative to the baseline EoS band constrained by the current NICER and GW170817 observations. The black contours denote the existing NICER and GW170817 observational constraints, while the cyan contours mark the optimal target selected in the previous iteration and are shown for reference.}
\label{fig:3R_01M_grid_heatmap}
\end{figure*}

Having shown that the cINN can map broad observational uncertainty, we systematically scan the $M$-$R$ phase space to identify an ideal sequence of six new hypothetical targets. Our objective is to determine the ordered sequence of new pulsar observations that maximally shrinks the EoS band inferred from current NICER and GW observations. At each step, we select the single best target conditioned on all prior observations to maximally narrow down current EoS uncertainties.

\begin{figure}[h]
    \centering
    \includegraphics[width=\linewidth]{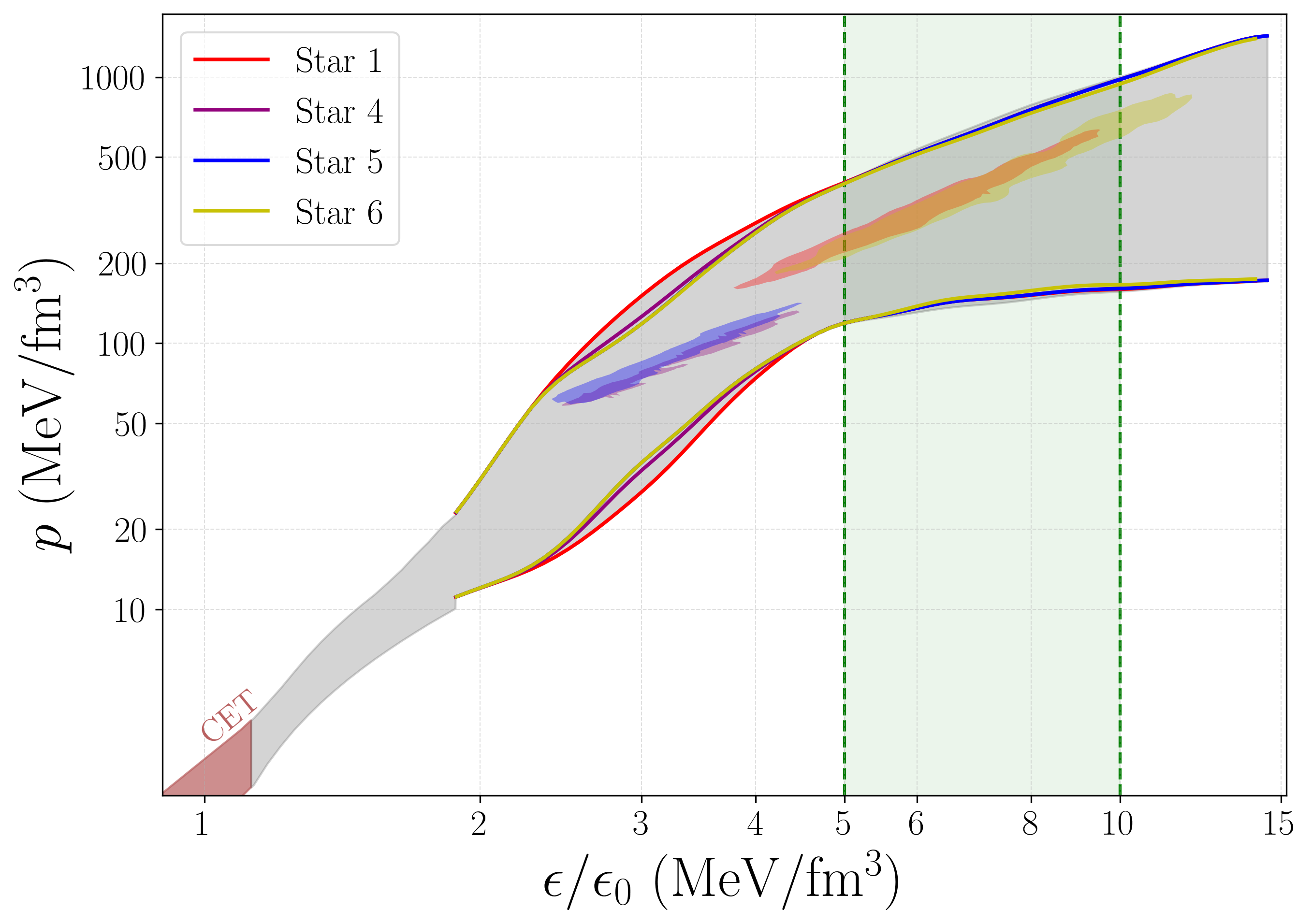}
    \caption{Iterative shrinkage of the predicted EoS band from sequential search. The grey envelope represents baseline observational uncertainty, while colored lines depict sequentially tightened constraints from optimal synthetic targets. Vertical green column indicates the central EoS values of maximum mass stars for monotonic and non-monotonic compositions. Colored blobs show cINN posteriors of stars exhibiting maximal EoS band shrinkage per iteration.}
    \label{fig:band_shrink_3R_01M}
\end{figure}

\begin{table}[h]
\centering
\caption{Stellar Parameters and Baseline Shrinkage Data}
\label{tab:shrinkage_3R_01M}
\resizebox{\linewidth}{!}{%
\begin{tabular}{lccccc}
\hline
\textbf{Star} & \textbf{Mass} & \textbf{Radius} & \multicolumn{2}{c}{\textbf{Shrinkage (dex)}} & \textbf{Percentage} \\
\cline{4-5}
\textbf{Number} & $\mathbf{(M)}$ & $\mathbf{(R)}$ & \textbf{wrt Prev} & \textbf{wrt Orig} & \textbf{wrt Orig (\%)} \\
\hline
1 & 2.04 & 11.00 & 0.0094 & 0.0094 & 1.60 \\
2 & 1.62 & 12.50 & 0.0082 & 0.0176 & 3.00 \\
3 & 1.54 & 12.30 & 0.0086 & 0.0262 & 4.47 \\
4 & 1.53 & 12.20 & 0.0078 & 0.0341 & 5.80 \\
5 & 1.63 & 12.50 & 0.0080 & 0.0420 & 7.15 \\
6 & 1.99 & 10.60 & 0.0079 & 0.0499 & 8.50 \\
\hline
\end{tabular}%
}
\end{table}

To quantify this shrinkage, we first establish a baseline EoS band using models that successfully satisfy current NICER observations. We then divide the $M$-$R$ plane into a high-resolution grid spanning $R \in [8.5, 16.5]$ km and $M \in [1.2, 2.5]\,M_\odot$. With step sizes of $\Delta R = 0.1$ km and $\Delta M = 0.01\,M_\odot$, this yields 10,400 distinct theoretical targets. We use a step-by-step search to find the optimal sequence of six stars. In each iteration, we scan all 10,400 grid points by evaluating them through the trained cINN ensemble. We identify the single target that most significantly reduces the current EoS uncertainty. Its synthetic data is then added to our baseline. The search continues iteratively: the new, narrower EoS band acts as the updated baseline-simulating the successful observation of the previous target and the full grid search is repeated to find the next most valuable target. 

The goal is to fully explore the $M$-$R$ space to find the maximum reduction in EoS uncertainty in six iterations, assuming these six new stars are observed alongside current NICER targets. This exercise highlights exactly which $M$-$R$ regions future X-ray telescope campaigns should focus on, rather than relying solely on improving overall measurement precision.
We quantify the reduction in EoS uncertainty by dividing the energy density into continuous bins. We define the local width, $W_i$, as the difference between the 10th and 90th percentiles of the predicted pressure distribution ($W_i = p_{90,i} - p_{10,i}$). The absolute information gain is the reduction in the mean band width, $\Delta W = \bar{W}_{\text{base}} - \bar{W}_{\text{joint}}$. Scanning the full contour set for six target stars required passing 62,400 distinct simulated observations through our model.

For the simulated observations, we assume measurement uncertainties of $0.1,M_\odot$ in mass and 3 km in radius, centred at every point of the $M$--$R$ grid. A sequential search is then performed to identify the observations that maximize the reduction of the EoS uncertainty. The resulting heatmaps are shown in Fig.~\ref{fig:3R_01M_grid_heatmap}, where each panel displays the incremental reduction in the EoS band relative to the baseline constrained by the current NICER and GW170817 observations. The black contours denote the existing observational constraints, while the cyan contours indicate the targets selected in previous iterations. For visual clarity, targets that lie very close to previously selected locations are omitted, although their positions are still indicated by the cyan contours. The corresponding reduction achieved at each iteration is summarized in Table~\ref{tab:shrinkage_3R_01M}.

Figure~\ref{fig:band_shrink_3R_01M} illustrates how the EoS envelope progressively contracts as the six optimal targets are incorporated. The red band denotes the chiral effective theory (CET) constraint, while our analysis focuses on densities above approximately $1.87\epsilon_0$ ($\epsilon_0\simeq150,\mathrm{MeV,fm^{-3}}$); lower densities are shown only for reference. The cINN-generated posteriors in the $(p_c,\epsilon_c)$ plane are colour-coded consistently with the selected targets in Fig.~\ref{fig:3R_01M_grid_heatmap}.

The sequential optimization reveals that the largest reduction in EoS uncertainty occurs at intermediate densities, approximately $2.5$--$4,\epsilon_0$, where the pressure predominantly determines the radii of intermediate- and high-mass neutron stars. Smaller improvements are also observed at higher densities ($6$--$10,\epsilon_0$), corresponding to the cores of maximum-mass stars populated by both monotonic and non-monotonic EoSs \cite{Verma:2025dez}. Notably, the optimization does not preferentially constrain the narrow high-density region associated with the onset of these maximum-mass configurations, indicating that present observational precision is primarily sensitive to the EoS at moderate supranuclear densities.
Starting from the baseline EoS width of 0.5877 dex, each newly selected target is incorporated sequentially to update the posterior before selecting the next observation. After six iterations, the EoS width decreases to 0.5377 dex, corresponding to an absolute reduction of 0.0499 dex, or an $8.5\%$ improvement over the baseline uncertainty.

In summary, our sequential search reveals highly specific, optimal target zones for future observational campaigns. We conducted a similar study using a larger measurement uncertainty contour ($0.3\,M_\odot$ and 2 km) in Appendix \ref{app:contour_2R_03M}. The optimal $M$-$R$ targets remained largely unchanged, indicating that changes in detector sensitivity do not significantly affect the overall strategy for constraining the EoS.

\section{Summary and Conclusion}
\label{sec:summary}

We have demonstrated that diffuse observational mass--radius ($M$--$R$) contours, such as those obtained from NICER and gravitational-wave observations, can be mapped directly to the corresponding central energy density and pressure using a hybrid conditional Invertible Neural Network with Physics-Informed Regularisation (cINN+PIR). Unlike conventional Bayesian inference, this framework bypasses forward modelling entirely, eliminating the need to construct parameterised equations of state (EoSs) and repeatedly solve the Tolman--Oppenheimer--Volkoff equations during inference. The resulting inverse mapping is performed within seconds while faithfully propagating observational uncertainties into the high-density EoS parameter space.

The PIR framework ensures that all inferred solutions satisfy fundamental physical requirements, enforcing both causality ($0 \le c_s^2 \le 1$) and thermodynamic stability. Tests on synthetic NICER-like observations demonstrate that the model accurately recovers distinct central-density and central-pressure posteriors while preserving the underlying relationship between the macroscopic stellar observables and the microscopic properties of dense matter.

Leveraging the computational efficiency of the framework, we performed a systematic optimization study involving 62,400 simulated observations. This analysis shows that the constraining power of future measurements depends strongly on their location in the mass--radius plane rather than solely on measurement precision. The optimal observing strategy alternates between compact high-mass stars and extended intermediate-mass stars, reducing the uncertainty in the inferred EoS by up to $8.5\%$ relative to the current NICER baseline. These results provide concrete observational guidance for future X-ray and multi-messenger campaigns by identifying the most informative regions of the mass--radius parameter space.

A comparison with Bayesian MCMC inference demonstrates that the two approaches are complementary. While MCMC provides sharper localization around the highest-likelihood solutions, the cINN offers substantially better global coverage of the physically allowed high-density EoS space and maintains robust performance even in extrapolated regions where conventional parametric sampling becomes inefficient.

Overall, this work establishes the cINN+PIR framework as a physically consistent, computationally efficient approach for solving the inverse neutron-star EoS problem. More broadly, it illustrates how incorporating exact physical constraints into deep generative models can transform astrophysical inverse inference, enabling rapid, reliable, and physically interpretable analyses for next-generation multi-messenger observations.

\section{Acknowledgments}
The authors thank the Indian Institute of Science Education and Research Bhopal for providing all the research and computation infrastructure facilities. The authors also acknowledge Dr. Anshuman Verma for providing crucial insights to this project. RM acknowledges the Science and Engineering Research Board (SERB), Govt. of India, for financial support in the form of Core Research Grant (CRG/2022/000663).

\section{Data Availability}
Model generated data corresponding to Figures \ref{fig:synthetic_contour}, \ref{fig:cs2_vs_E}, \ref{fig:cinn_blobs} and \ref{fig:band_shrink_3R_01M} are available on Zenodo repository \cite{utkarsh_deshmukh_2026_21337922}.

\section*{APPENDIX}

\makeatletter
\renewcommand{\p@subsection}{} 
\makeatother
% ----------------------------------------------------

\setcounter{subsection}{0}
\renewcommand{\thesubsection}{\Alph{subsection}}
% ----------------------------------------------------------------

\subsection{Visual Representation of Model Architecture}
\label{app:model_pipeline}

\begin{figure*} 
    \includegraphics[width=\linewidth]{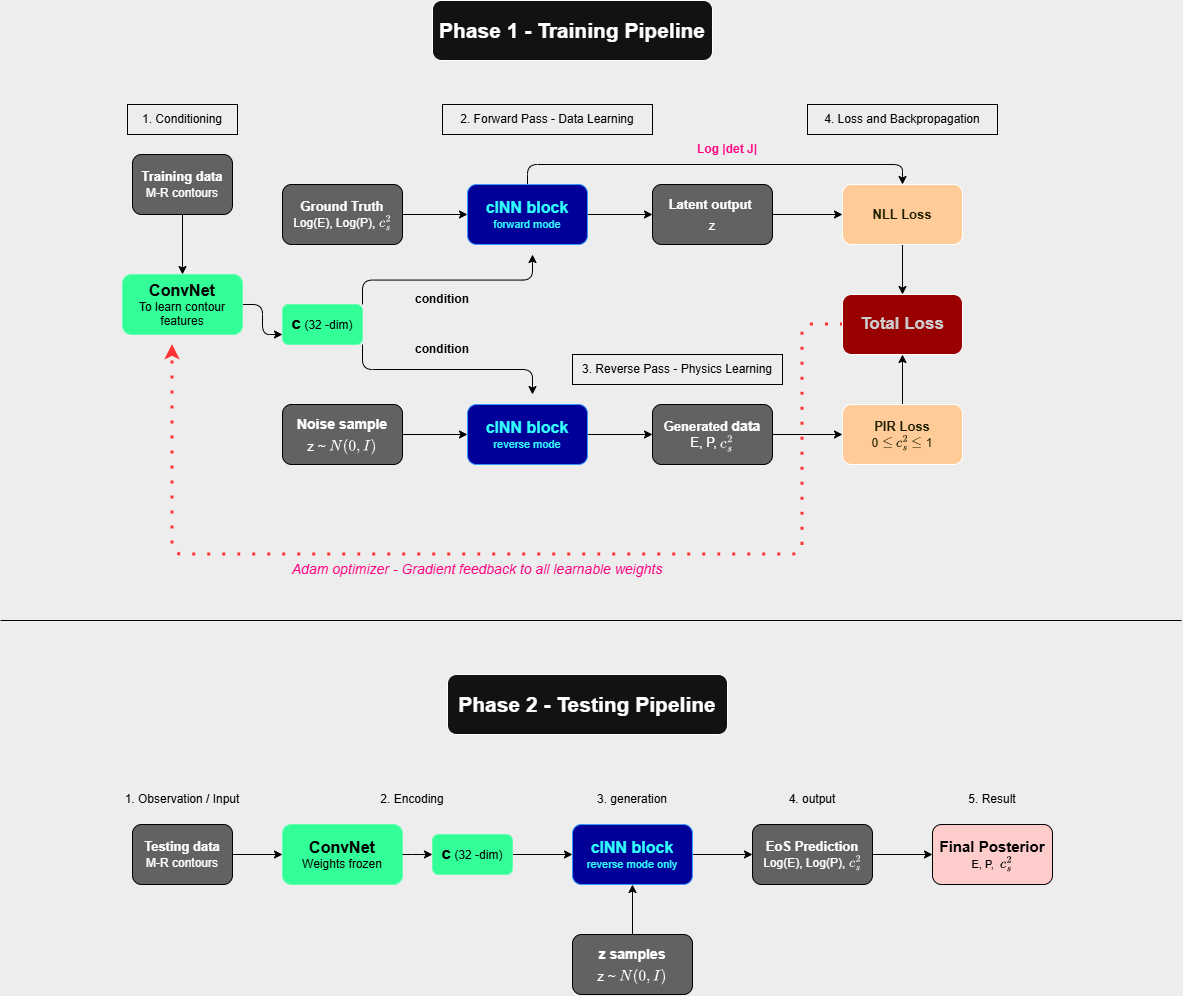}
     \caption{Architecture of the cINN-PIR model pipeline, illustrating the forward likelihood optimization and the backward physics-informed regularization.}
    \label{fig:cinn_model_pipeline}
\end{figure*}

The training and testing model pipeline for cINN+PIR architecture is visually detailed in Figure \ref{fig:cinn_model_pipeline}

% \onecolumngrid
% \vspace*{\fill} 

% \twocolumngrid

% ------------------------------------------------------------------
\subsection{Sensitivity Analysis: Impact of Varying Observational Uncertainties}
\label{app:contour_2R_03M}

\begin{figure*}[ht]
    \includegraphics[width=\linewidth]{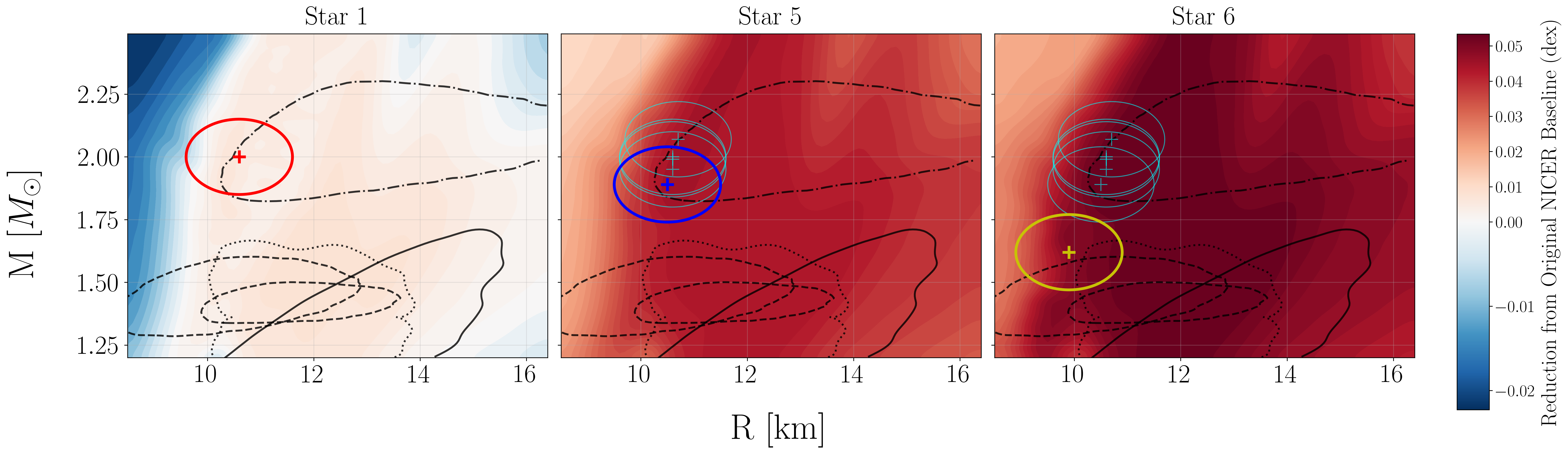}
    \caption{Evolution of 0.3$M_\odot$ and 2 km radius stars from the NICER baseline band shrinkage across the $M$-$R$ phase space during the sequential search for six optimal targets.}
    \label{fig:2R_03M_grid_heatmap}
\end{figure*}

\begin{figure}[h]
    \centering
    \includegraphics[width=\linewidth]{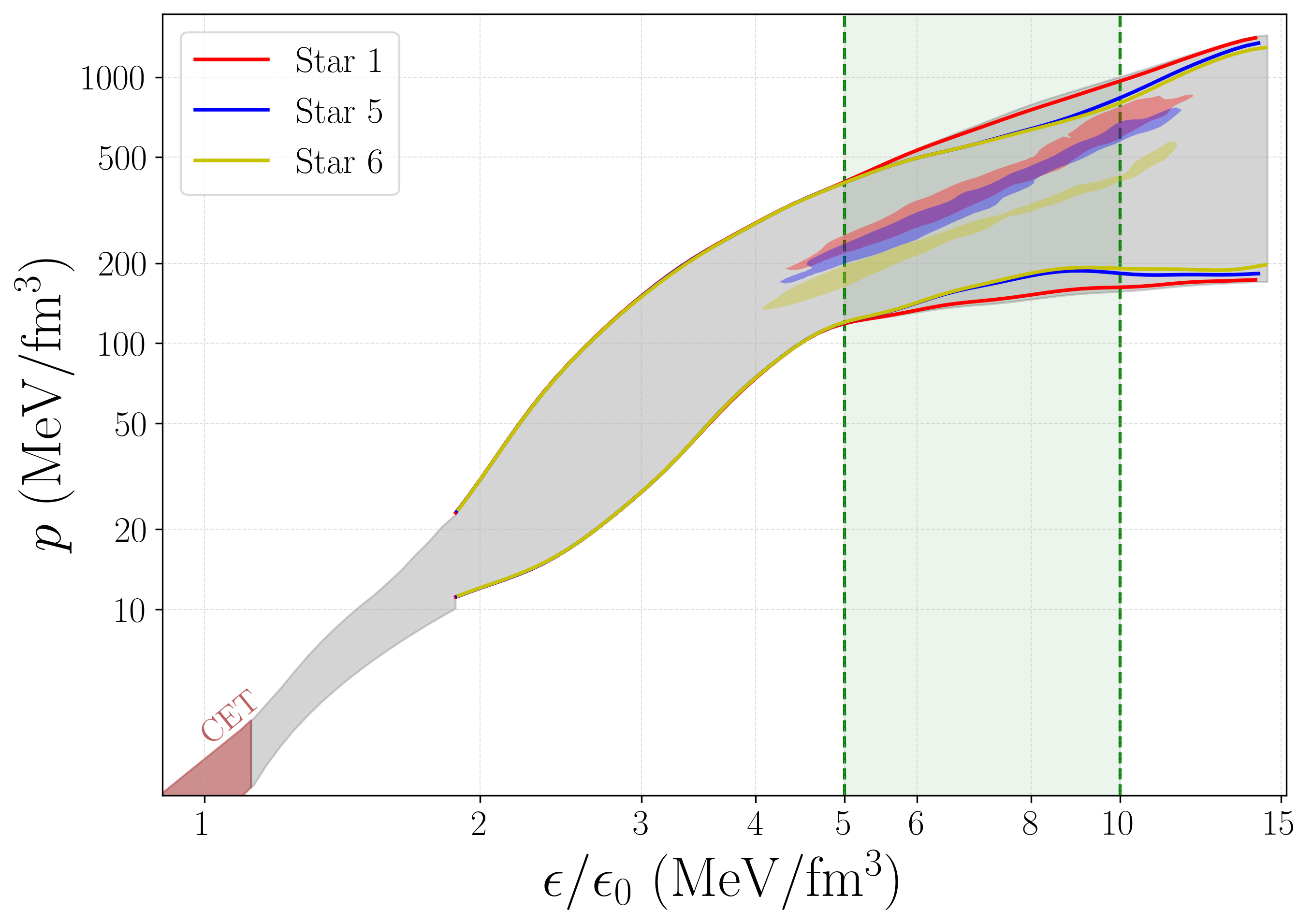}
    \caption{Iterative shrinkage of the predicted EoS band resulting from the  sequential search.}
    \label{fig:band_shrink_2R_03M}
\end{figure}

We conduct an analysis parallel to that in Section \ref{sec:heatmap_3R_01M}, but modify the simulated measurement uncertainties to $0.3\,M_\odot$ in mass and 2 km in radius. This tests whether variations in detector sensitivity significantly alter our conclusions regarding EoS constraints. Additionally, this exercise demonstrates the model's versatility in processing observational contours of varying sizes to generate high-density EoS posteriors. Due to visual proximity of stars 2, 3, 4 close to each other, only three stars at distant locations in the $M$-$R$ space are plotted in Figure \ref{fig:2R_03M_grid_heatmap} with their corresponding band shrinkage visualized in Figure \ref{fig:band_shrink_2R_03M}. The contours plotted in cyan represent the previous iteration's star. The progressive reductions in EoS uncertainty for each iteration are detailed in Table \ref{tab:shrinkage_2R_03M}.

The results indicate that broadening the observational uncertainty contours does not drastically alter the overall reduction in the EoS band width. Starting from an initial NICER baseline uncertainty of 0.5877 dex, the band narrows to 0.5321 dex by the end of the sixth iteration. This corresponds to a 9.46\% total reduction in uncertainty from the baseline. Compared to the 8.50\% shrinkage achieved with tighter contours, varying the $M$-$R$ detector sensitivity does not fundamentally alter the effectiveness of this targeting strategy. More importantly, this test highlights a practical limitation: saturating the $M$-$R$ phase space with broad, less precise observational contours yields diminishing returns when attempting to further constrain the EoS. 

\begin{table}[h]
    \centering
    \caption{Stellar Parameters and Baseline Shrinkage Data}
    \label{tab:shrinkage_2R_03M}
    \resizebox{\linewidth}{!}{%
    \begin{tabular}{lccccc}
    \hline
    \textbf{Star} & \textbf{Mass} & \textbf{Radius} & \multicolumn{2}{c}{\textbf{Shrinkage (dex)}} & \textbf{Percentage} \\
    \cline{4-5}
    \textbf{Number} & $\mathbf{(M)}$ & $\mathbf{(R)}$ & \textbf{wrt Prev} & \textbf{wrt Orig} & \textbf{wrt Orig (\%)} \\
    \hline
    1 & 2.00 & 10.60 & 0.0094 & 0.0094 & 1.59 \\
    2 & 1.95 & 10.60 & 0.0092 & 0.0185 & 3.15 \\
    3 & 1.99 & 10.60 & 0.0088 & 0.0273 & 4.65 \\
    4 & 2.07 & 10.70 & 0.0088 & 0.0362 & 6.15 \\
    5 & 1.89 & 10.50 & 0.0093 & 0.0455 & 7.74 \\
    6 & 1.62 & 9.90 & 0.0101 & 0.0556 & 9.46 \\
    \hline
\end{tabular}%
}
\end{table}

\subsection{Comparing with Bayesian Analysis}
\label{app:bayesian_comparison}

We compare our work with Bayesian EoS reconstruction. This method models neutron star interiors using the SoSI approach with parameters $\gamma_{CET}$, $c_1$, $c_2$, $c_3$, and $c_4$ (see section \ref{sec:data}). It enforces strict physical rules: causality ($c_{s}^2 < 1$) and stability ($c_{s}^2 > 0$). MCMC algorithms sample variables like $c_{s}^2$ and $\Gamma_{\text{BPS}}$ to find EoS setups that match observed $M$-$R$ curves. While reliable and physically sound, the Bayesian approach has built-in traits that limit its output being a forward model. It rejects any setup breaking physical rules, cutting off parts of the 95\% $M$-$R$ contour that are statistically possible but physically unreachable. Second, MCMC favors the center of the data space where curves are easiest to generate, often ignoring the extreme edges of the observations. Therefore, Bayesian results show a very safe but restricted portion of the data, missing the full range of uncertainty. In contrast, our framework works as a continuous inverse-mapper. Instead of pushing fixed values through the TOV equations, the cINN maps the entire given input $M$-$R$ space directly back to it's corresponding central EoS space ($\epsilon_c$, $p_c$). To keep results physically valid, we use PIR. The cINN flexibly maps the data's shape, while PIR acts as a strict rule, penalizing paths that break causality or stability. This combined cINN+PIR method quickly maps the outer limits of the allowed data space. It offers a fast, thorough alternative to Bayesian analysis while still obeying physical laws. 

\begin{figure}[h]
    \centering
    \includegraphics[width=\linewidth]{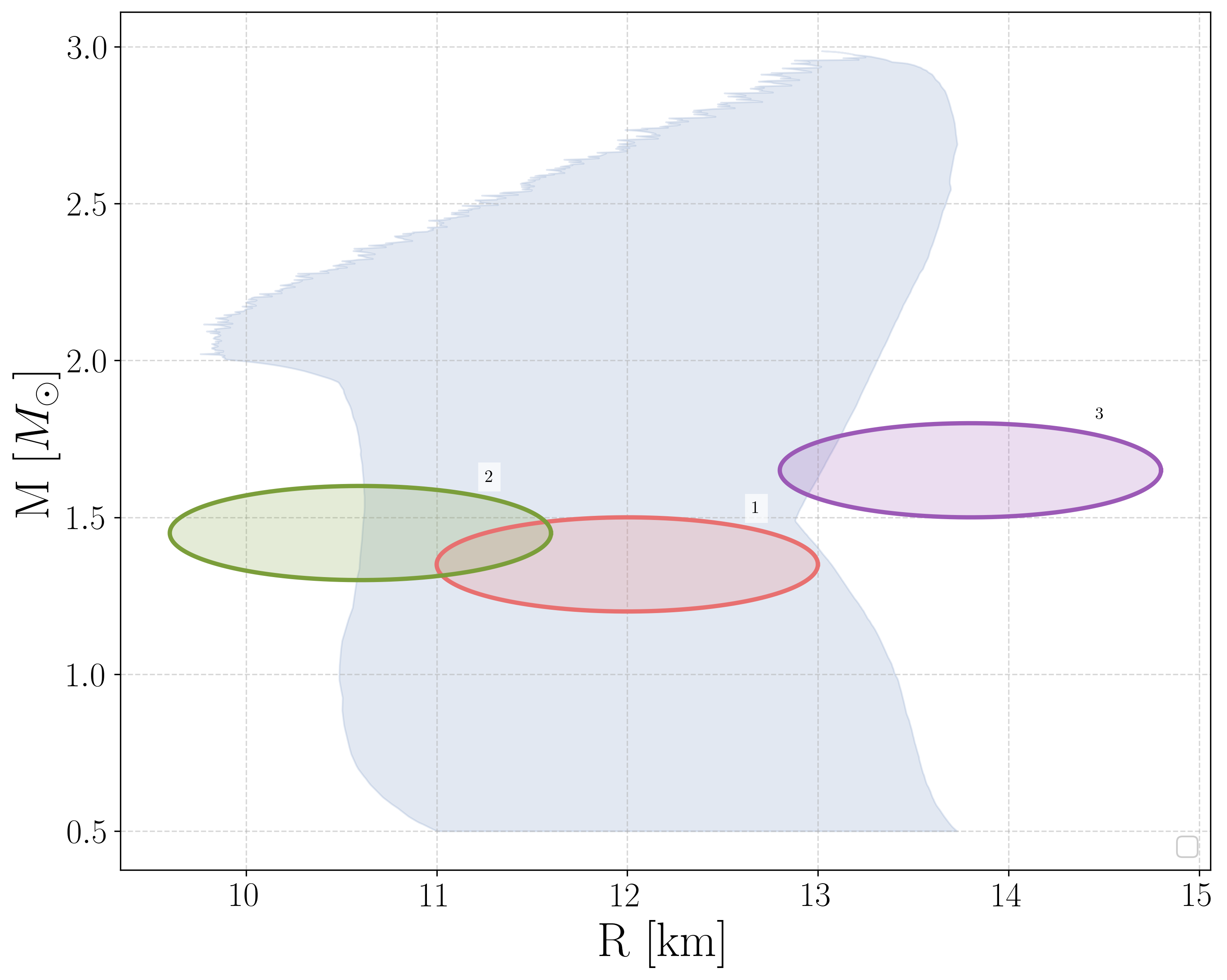}
    \includegraphics[width=\linewidth]{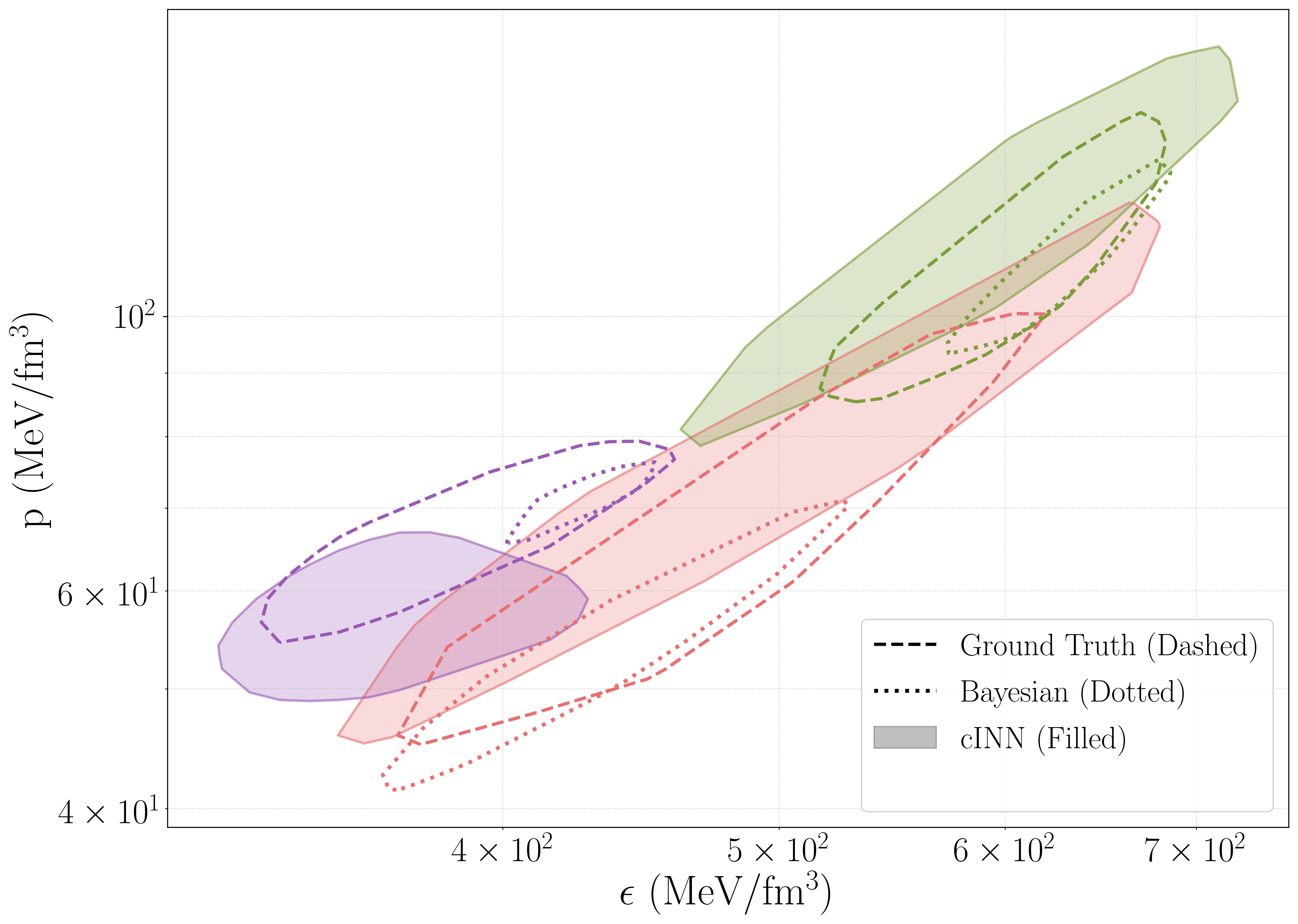}
    \caption{\textit{Upper panel :-} Synthetic mass radius contours annotated based on their overlap with training data range. \textit{Lower panel :-} Comparison plot showing central value contours for the three synthetic stars derived from different methods as labeled. Plotted in logarithmic scale, each color of contour corresponds to the respective annotated star in Fig. \ref{fig:6} (upper panel).}
\label{fig:6}
\end{figure}

To test our model, we use synthetic $M$-$R$ contours with fixed sizes ($\Delta M = 0.3 M_\odot$ and $\Delta R = 2$ km). We designed these to have different amounts of overlap with our Ground Truth (GT) training data. The GT includes various EoS curves created by solving TOV equations. By changing the overlap, we see how much data falls in the "known" range versus the "unknown" range where the model must guess (extrapolate). This checks how well the cINN performs when training data is sparse. We compare the network's predictions for $\epsilon_c$ and $p_c$ against the GT and Bayesian results to verify its accuracy in unknown areas. Matching the Bayesian reliability proves our cINN+PIR model is a fast and trustworthy alternative. We use three specific $M$-$R$ contours based on their overlap, shown in Figure \ref{fig:6} (upper panel):

\begin{itemize}
\item \textbf{Star 1:} 99\% overlap with the training data range.
\item \textbf{Star 2:} 51.91\% overlap with the training data range.
\item \textbf{Star 3:} 15.30\% overlap with the training data range.
\end{itemize}

Figure \ref{fig:6} (lower panel) shows the outer boundaries in the $\epsilon_c$-$p_c$ space, comparing the GT (dashed lines), Bayesian results (dotted lines), and cINN predictions (filled regions) for the three stars. The Bayesian results cover a much smaller, tighter area compared to the GT and cINN. This happens because parametric EoS models naturally narrow down. By enforcing causality ($c_{s}^2 < 1$) and stability ($c_{s}^2 \ge 0$), the Bayesian SoSI method guarantees safe, physical results. However, it loses flexibility. Its fixed shape cannot easily stretch to cover the outer edges of a 95\% confidence $M$-$R$ contour. While this makes Bayesian outputs very precise, using them alone can hide the true size of the observational uncertainty.

\begin{figure}[h]
    \centering
    \includegraphics[width=0.8\linewidth]{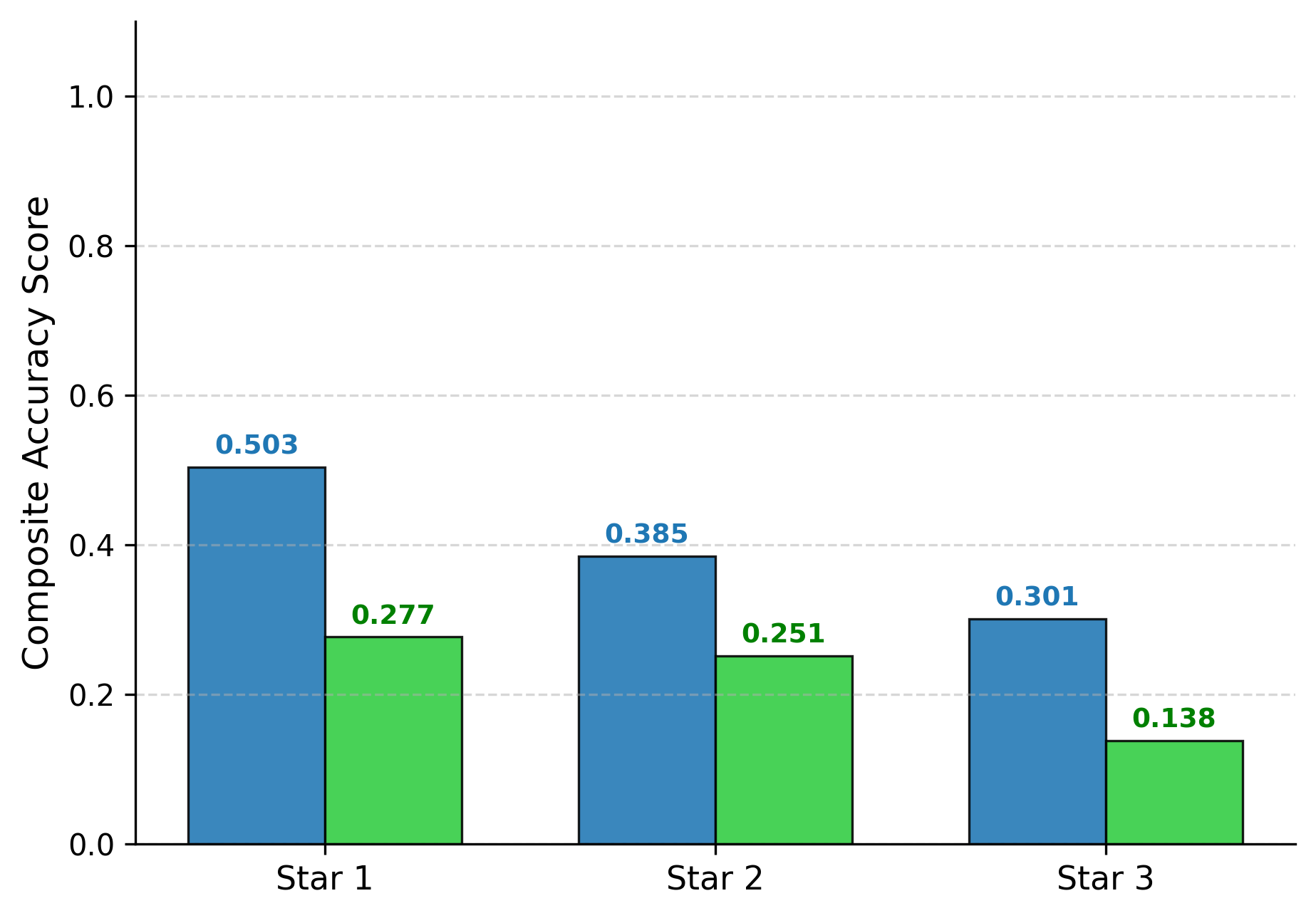}
    \caption{Composite accuracy score (Mean of Jaccard and Dice indices) comparing the cINN and Bayesian models across the three synthetic star scenarios. Higher values indicate better overall spatial agreement with the GT.}
    \label{fig:error_metric_3}
\end{figure}

In contrast, the cINN maps data backward continuously without relying on a fixed model. Instead of fitting a rigid curve, it translates the full $M$-$R$ contours directly into the $\epsilon_c$-$p_c$ space. As seen in Figure \ref{fig:6} (lower panel), the cINN smoothly covers the GT, filling in valid physical areas missed by limited datasets. Star 3 is a key test of the model's ability to extrapolate. With only 15.30\% of its data in the training range, the cINN naturally shifts to lower central energy densities and pressures. This shift is not random, rather, it makes physical sense since star 3 has low compactness (mass $M \le 1.75 M_\odot$ and radius $R \ge 12.5$ km). The network follows the correct TOV equation trends, pushing the central state toward the lower limits of the EoS space. This proves the cINN learned the actual physics, rather than just memorizing data, allowing it to make valid predictions even where data is lacking. To measure how much the cINN and Bayesian results differ from the GT, we use comparative metrics. We check the outer edges of each predicted EoS region in the $\log_{10}\epsilon_c$ - $\log_{10}p_c$ space to find overlaps, shape differences, and bias. All distances here are in $\log_{10}$ units (dex). We measure spatial overlap using the Jaccard Index (intersection divided by union) and the Dice Coefficient (twice the intersection divided by total area). These penalize both missing valid areas and guessing too far outside, giving a balanced score. We combine these into a single composite accuracy score (their average). As illustrated in Figure \ref{fig:error_metric_3}, this composite score confirms that the cINN consistently beats the Bayesian method at capturing the valid GT area across all cases.

\begin{figure}[htbp]
\centering
\includegraphics[width=\linewidth]{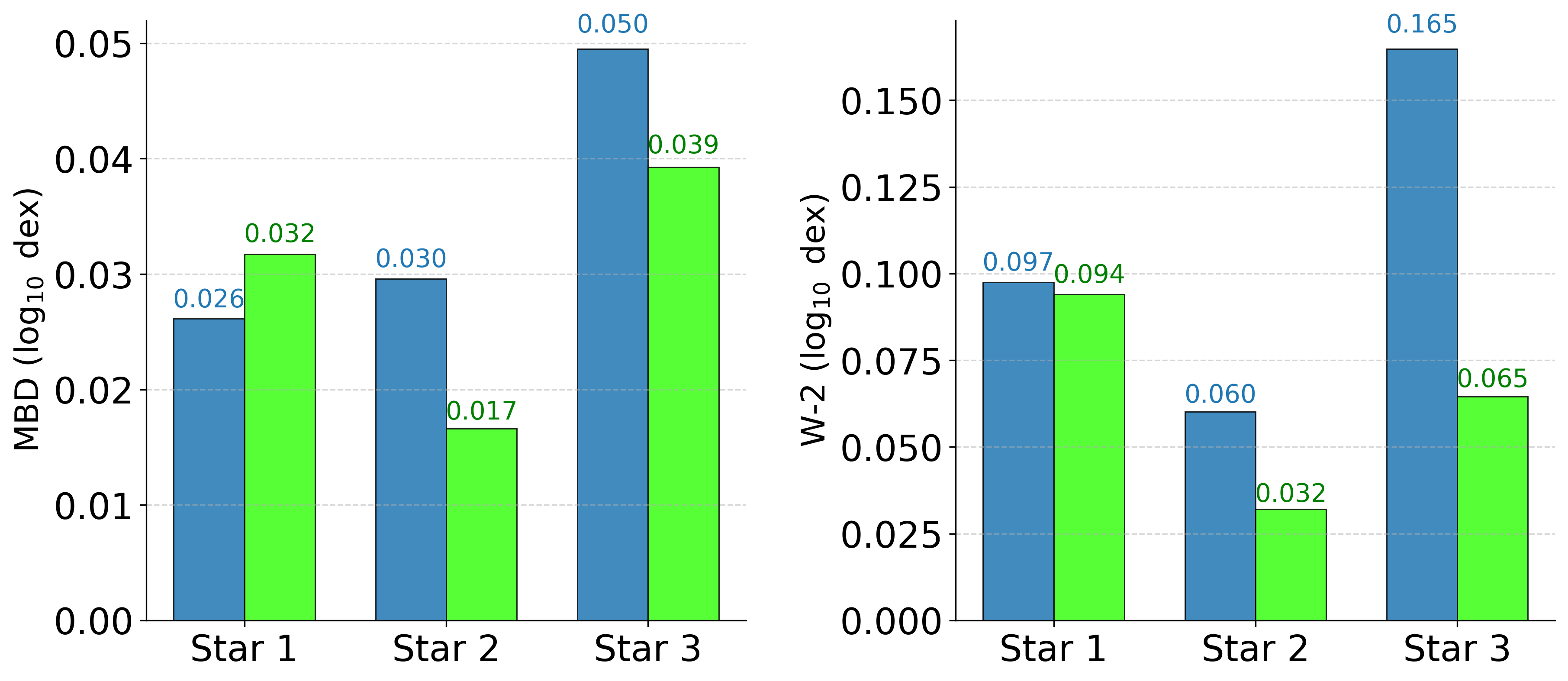}
\caption{Error metrics comparing the cINN (blue) and Bayesian (green) models. \textit{Left:} Mean Boundary Distance (lower indicates a tighter average boundary). \textit{Right:} Wasserstein-2 Distance (lower indicates better probability mass alignment).}
\label{fig:error_metric}
\end{figure}

Because the Bayesian method strictly follows fixed bounds to ensure physical safety, it focuses on a small, high-confidence area. It covers only 30.6\%, 20.8\%, and 9.8\% of the GT area for Stars 1, 2, and 3 respectively. In contrast, the cINN maps a much wider area, reaching ratios of 1.28 and 1.46 for Stars 1 and 2. Even when forced to guess outside its main training data (Star 3), the cINN keeps a 67.1\% coverage, a composite score of 0.301 (versus the Bayesian 0.138), and a Jaccard Index of 0.2291 (versus 0.0980). This shows the cINN's main strength: quickly and fully mapping the wide observational uncertainty that traditional solvers struggle to reach. We also evaluate the models using Mean Boundary Distance (MBD) and Wasserstein-2 distance (W-2), shown in Figure \ref{fig:error_metric}. MBD measures the average distance between the predicted outer boundary and the true GT boundary.

For Star 1 (with plenty of training data), the cINN traces the true boundary better than the Bayesian model (0.026 vs. 0.032 dex). However, for Stars 2 and 3, the Bayesian model gets a lower (better) MBD. As mentioned before, this is because the Bayesian model is conservative; it shrinks to a small, safe area near the center, keeping its distance errors low. The cINN does the opposite: it tries to map the wider, unknown edges, which naturally increases its average geometric error. The W-2 score measures the mathematical "cost" to shift the predicted results to perfectly match the true data. For Star 1, both do well (0.097 vs. 0.094 dex). For Stars 2 and 3, the Bayesian model scores better because it packs its predictions tightly over the safest core of the GT, making its "moving cost" low. The cINN gets higher (worse) scores because it spreads out to capture the full uncertainty, which increases the average cost. Ultimately, these metrics show each method's role. The Bayesian method is precise and safe, finding the dense core. The cINN is exploratory, accepting slightly higher mathematical errors to successfully map the wide, true boundaries that traditional models miss. Essentially, while Bayesian MCMC remains an excellent tool for rigorous inference, heavily over-constraining the $p_c$-$\epsilon_c$ contours is not the ultimate goal of this project. It is unwise to assume that those regions are physically invalid given a traditional sampler struggles to reach them-like the outer edges of a NICER contour. Physically viable EoS configurations that satisfy observational data can and do exist in these peripheral spaces. The cINN directly addresses these blind spots. By trading a minor degree of localized precision for immense computational speed, the network overcomes traditional sampling bottlenecks. It recovers the peripheral, valid regions, providing a broader and much more complete mapping of the true observational uncertainty.

\twocolumngrid
\makeatletter
\def\bibsection{\section*{References}}
\makeatother
\bibliographystyle{unsrturl}
\bibliography{mybib}

\end{document}